\documentclass[preprint]{aastex631}

\def\gtrsim{\mathrel{\hbox{\rlap{\hbox{\lower5pt\hbox{$\sim$}}}\hbox{$>$}}}}
\def\lesssim{\lower.5ex\hbox{$\; \buildrel < \over \sim \;$}}
\newcommand{\msun}{\mbox{M$_\odot$}}

\newcommand{\kms}{km s$^{-1}$}

\newcommand{\mjybeam}{mJy~beam$^{-1}$}

\begin{document}

\title{Triple spiral arms of a triple protostar system imaged in molecular lines}

\correspondingauthor{Jeong-Eun Lee}
\email{lee.jeongeun@snu.ac.kr}

\author[0000-0003-3119-2087]{Jeong-Eun Lee}
\affil{Department of Physics and Astronomy, Seoul National University \\
1 Gwanak-ro, Gwanak-gu, Seoul 08826, Korea}

\author{Tomoaki Matsumoto}
\affiliation{Faculty of Sustainability Studies, Hosei University \\
Fujimi, Chiyoda-ku, Tokyo, 102-8160 Japan}

\author[0000-0001-9263-3275]{Hyun-Jeong Kim}
\affiliation{Korea Astronomy and Space Science Institute \\
776 Daedeok-daero, Yuseong-gu, Daejeon, 34055, Korea}

\author{Seokho Lee}
\affiliation{Korea Astronomy and Space Science Institute \\
776 Daedeok-daero, Yuseong-gu, Daejeon, 34055, Korea}

\author{Daniel Harsono}
\affiliation{Institute of Astronomy and Department of Physics, National Tsing Hua University \\
Hsinchu 30013, Taiwan}

\author{Jaehan Bae}
\affiliation{Department of Astronomy, University of Florida \\
Gainesville, FL 32611, USA}

\author{Neal J. Evans II}
\affiliation{Department of Astronomy, The University of Texas at Austin \\
515 Speedway, Stop C1400, Austin, TX 78712-1205, USA}

\author{Shu-ichiro Inutsuka}
\affiliation{Department of Physics, Nagoya University \\
Furo-cho, Chikusa-ku, Nagoya, Aichi 464-8602, Japan}

\author{Minho Choi}
\affiliation{Korea Astronomy and Space Science Institute \\
776 Daedeok-daero, Yuseong-gu, Daejeon, 34055, Korea}

\author{Ken'ichi Tatematsu}
\affiliation{National Astronomical Observatory in Japan \\
2-21-1 Osawa, Mitaka, Tokyo 181-8588, Japan}

\author{Jae-Joon Lee}
\affiliation{Korea Astronomy and Space Science Institute \\
776 Daedeok-daero, Yuseong-gu, Daejeon, 34055, Korea}

\author{Dan Jaffe}
\affiliation{Department of Astronomy, The University of Texas at Austin \\
515 Speedway, Stop C1400, Austin, TX 78712-1205, USA}



\begin{abstract}

Most stars form in multiple star systems. For a better understanding of their formation processes, it is important to resolve the individual protostellar components and the surrounding envelope and disk material at the earliest possible formation epoch because the formation history can be lost in a few orbital timescales. 
Here we present the ALMA observational results of a young multiple protostellar system, IRAS 04239+2436, where three well developed large spiral arms were detected in the shocked SO emission. Along the most conspicuous arm, the accretion streamer was also detected in the SO$_2$ emission.
The observational results are complemented by numerical magneto-hydrodynamic simulations, where those large arms only appear in magnetically weakened clouds. The numerical simulations also suggest that the large triple spiral arms are the result of gravitational interactions between compact triple protostars and the turbulent infalling envelope.

\end{abstract}


\section{Introduction}\label{sec:intro}

Although binary or multiple formation is the main mode of star formation \citep{offner22}, their main formation mechanism remains elusive. 
Observational and theoretical studies that zoom into the young circumstellar and circumbinary disks in the early evolutionary stage is critical for understanding the formation process before their history is erased through dynamical interactions \citep{Moe2018, Lee2017}. Interactions between the inner envelope and the central protostellar system, as well as the interactions among individual members, are important in multiple formation since protostars grow their masses the most aggressively in the early formation stage when most of mass still remains in the envelope. Therefore, simultaneous imaging of the inner envelope and the central compact multiple system is also required. 
One prominent structure that shows the interactions between forming binary and circumbinary material is spiral arm structures \citep{Alves2019, Takakuwa2017}.

Two distinct formation mechanisms of multiple star systems have been suggested; turbulent fragmentation for wide ($>500$ au) binaries \citep{Offner2010,Lee2017} and disk fragmentation for close ($<500$ au) binaries \citep{kratter2010, Alves2019}. However, at early stages, when the envelope is feeding material to the disk(s), the clean distinction breaks down.
Instead, a hybrid process, where the turbulent envelope plays an important role in the disk fragmentation, the circumbinary/circumstellar disk structures, and the final masses of the individual central sources, may be important \citep{Matsumoto2015}.
In addition to turbulence, the magnetic field must also play an important role in the fragmentation process itself and the structures inside fragments since cores with weak magnetic fields tend to form multiple systems \citep{Machida2005}. 


To study the formation process of a close binary system when the environmental effect of the natal cloud may still remain in the system, we observed IRAS 04239+2436 with the Atacama Large Millimeter/submillimeter Array (ALMA) simultaneously in the 857 $\mu$m continuum and several molecular lines, with the angular resolution of $\sim$0.1\arcsec\ ($\sim$14~au at 140~pc).
IRAS 04239+2436 is 
an embedded close protobinary with the projected separation of 42~au at a distance of 140~pc \citep{Reipurth2000}. 
IRAS 04239+2436 has a luminosity similar to that of the Sun and very rich near-infrared (NIR) spectrum \citep{Greene1996}. A high-spectral resolution NIR observation kinematically revealed a sub-au scale gaseous disk with a Keplerian rotation around the central mass of 0.14$\sim$0.2 \msun\  \citep{Lee2016}. 
In the NIR images (2~$\mu$m) of IRAS 04239+2436 \citep{Reipurth2000}, 
the eastern source (Source A) is about 2.65 times brighter than the western source (Source B), and is thus considered as the primary component. 
In addition, a well collimated northeast [Fe~\textsc{ii}] 1.644~$\mu$m jet of HH 300 originates from Source A \citep{Reipurth2000}. Yet, the observed jet precesses at a short timescale ($<$ 30 yr),
compared to the orbital period ($\sim$270 yr) anticipated from the resolved binary, indicative of an unresolved very close binary interactions \citep{Reipurth2000}. Therefore, IRAS 04239+2436 is very likely a triple system. 

Here, we report the ALMA observational results of a forming multiple protostellar system, IRAS 04239+2436 (Section \ref{sec:alma}) and its comparisons with numerical simulations (Section \ref{sec:simulation}).
We discuss the effect of magnetic field on multiple formation and the implication of close multiple protostars in planet formation in Section \ref{sec:discussion}, and the final conclusion is given in Section \ref{sec:conclusion}.

\section{ALMA Observations and Results}
\label{sec:alma}

IRAS 04239+2436 was observed using ALMA 
during  Cycle 3  (2015.1.00397.S, PI: Jeong-Eun Lee) on 2016 August 16 UT. 
Five spectral windows in Band 7 were set to cover several molecular lines such as SO 8$_8$--7$_7$ (344.31061200 GHz), HCN 4--3 (354.50547590 GHz), HCO$^+$ 4--3 (356.73422300 GHz), and SO$_2$ 10$_{4,6}$--10$_{3,7}$ (356.75518930 GHz).\footnote{The molecular line information is adopted from the Cologne Database for Molecular Spectroscopy \citep[CDMS;][]{CDMS} and  Jet Propulsion Laboratory \citep[JPL;][]{JPL} molecular databases.} 
The source velocity is $V_{\rm LSR} = 6.5$~\kms\  \citep{Fuller2002}.
The bandwidths and spectral resolutions were 117.19 MHz and 122.070 kHz ($\sim$0.1 \kms) for 
the SO line and 468.75 MHz and 244.141 kHz ($\sim$0.2 \kms) for the other three.
Forty 12-m antennas were used in the C40-4 configuration with baselines in the range from 
21.4 m to 3.1 km. 
The total observing time on source was 29.8 minutes. 

The data were initially calibrated using the CASA 4.7.0 pipeline \citep{McMullin2007}.
The nearby quasar J0510+1800 was used as a bandpass and phase calibrator, 
and the quasar J0238+1636 was used as an amplitude calibrator. 
Self-calibration was applied for better imaging.  The continuum image with the briggs weighting (robust=0.5) is used for the self-calibration. Two phase calibrations with ‘inf’ and ‘60s’ intervals are applied, which increases the S/N of the continuum image by a factor of 5.
The molecular line images were produced by the {\it clean} task within CASA with 
natural weighting. The synthesized beams are $0.184\arcsec \times 0.119\arcsec$, $0.201\arcsec \times 0.124\arcsec$, $0.180\arcsec \times 0.119\arcsec$, and $0.202\arcsec \times 0.125\arcsec$ respectively for SO, SO$_2$, HCN, and HCO$^+$ 
with the position angles (PAs) of $10.1\arcdeg$, $19.5\arcdeg$, $11.4\arcdeg$, and $19.5\arcdeg$.
The rms noise levels ($\sigma$) for the SO, SO$_2$, HCN, and HCO$^+$ images are 3.5, 4.45, 4.15, and 4.17 mJy beam$^{-1}$, respectively.
The moment maps of molecular lines were generated by the {\it immoments} task 
using the threshold of $6\sigma$. 
The continuum image was produced by using line-free channels with uniform 
weighting. The synthesized beam is $0.14\arcsec \times 0.08\arcsec$
with PA of $4.3\arcdeg$, and the rms noise level is 0.15 mJy beam$^{-1}$.

\begin{figure*}
\includegraphics[width=0.9\textwidth]{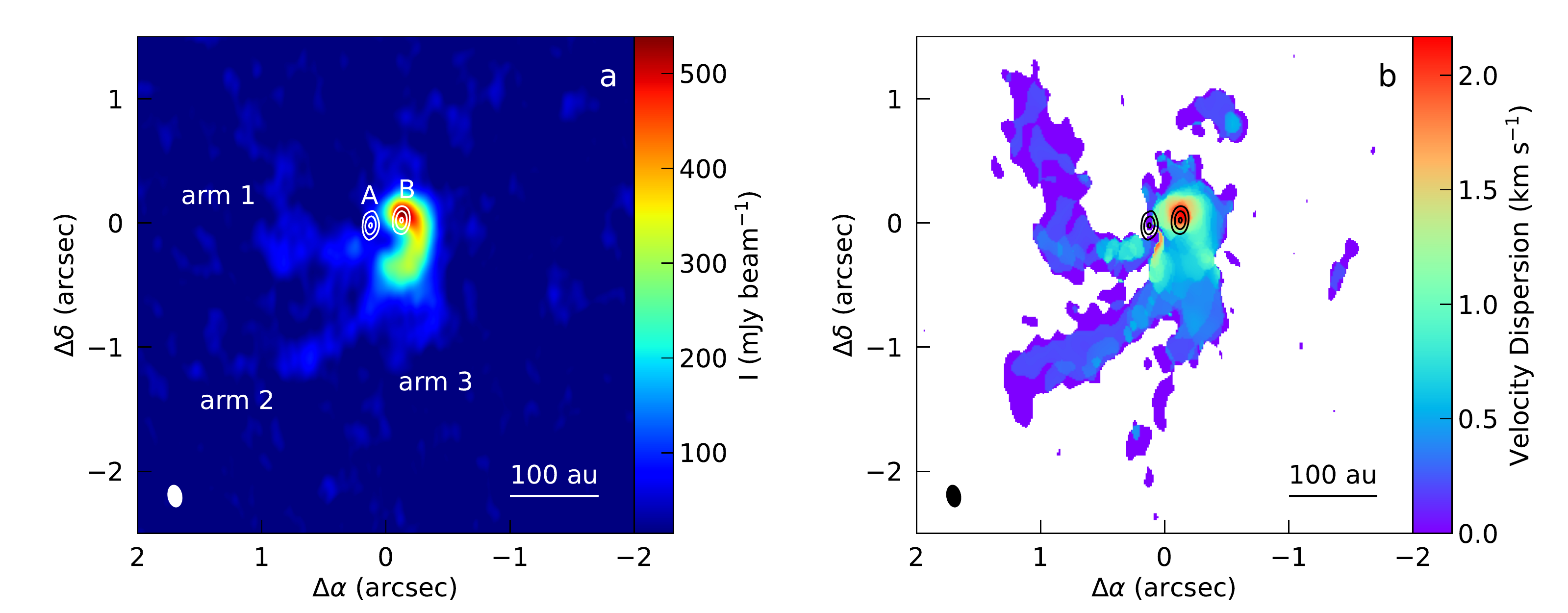}
	\caption{The SO integrated intensity (a) and velocity dispersion (b) maps.
	White (a) and black (b) contours depict the continuum emission distribution. The ellipse in the lower left corner represents the beam size. The contour levels and the beam size of the continuum image are the same in all figures.The SO integrated intensity map shows a strong flow-like structure toward Source B, and the velocity dispersion also increases toward Source B. The individual line profiles can be found in Fig.~\ref{fig_SO_specmap}; the SO lines around Source B is very broad, while the line intensity itself peaks at the flow-like structure beneath Source B.
	The integrated intensity is proportional to the column density multiplied by the excitation temperature in the molecular gas, and thus, it increases rapidly toward the central source. The peak intensity of the SO line is more sensitive to the excitation temperature of the molecule, so it traces better the shocked gas along the spiral arms, as presented in Fig.~\ref{fig_obs_simul}a.
} 
\label{fig_SO_mom02}
\end{figure*}

\begin{figure*}
\includegraphics[width=1.0\textwidth]{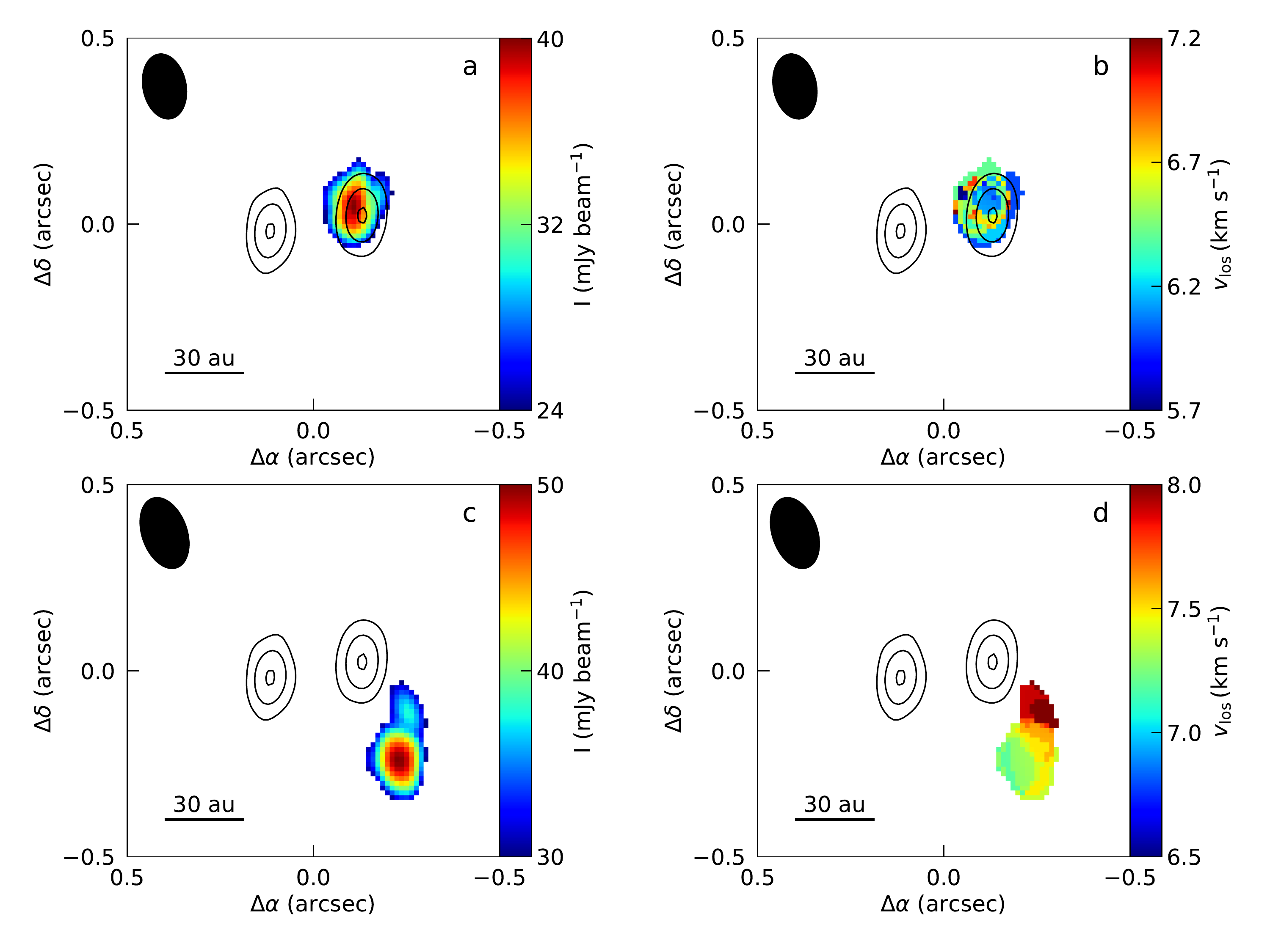}
	\caption{The peak intensity map and the intensity weighted velocity map of HCN J=4--3 (a, b) and SO$_2$ (c, d).
	The maps were generated using the threshold of $6\sigma$ (0.024 and 0.03 Jy beam$^{-1}$ for HCN and SO$_2$, respectively). The contours show the continuum emission at 857 $\mu$m. The HCN emission appears only in Source B. The SO$_2$ emission traces the accreting gas toward Source B.  The ellipse in the upper left corner represents the beam size. 
}
\label{fig_HCN_SO}
\end{figure*}

\begin{figure*}
\includegraphics[width=0.98\textwidth]{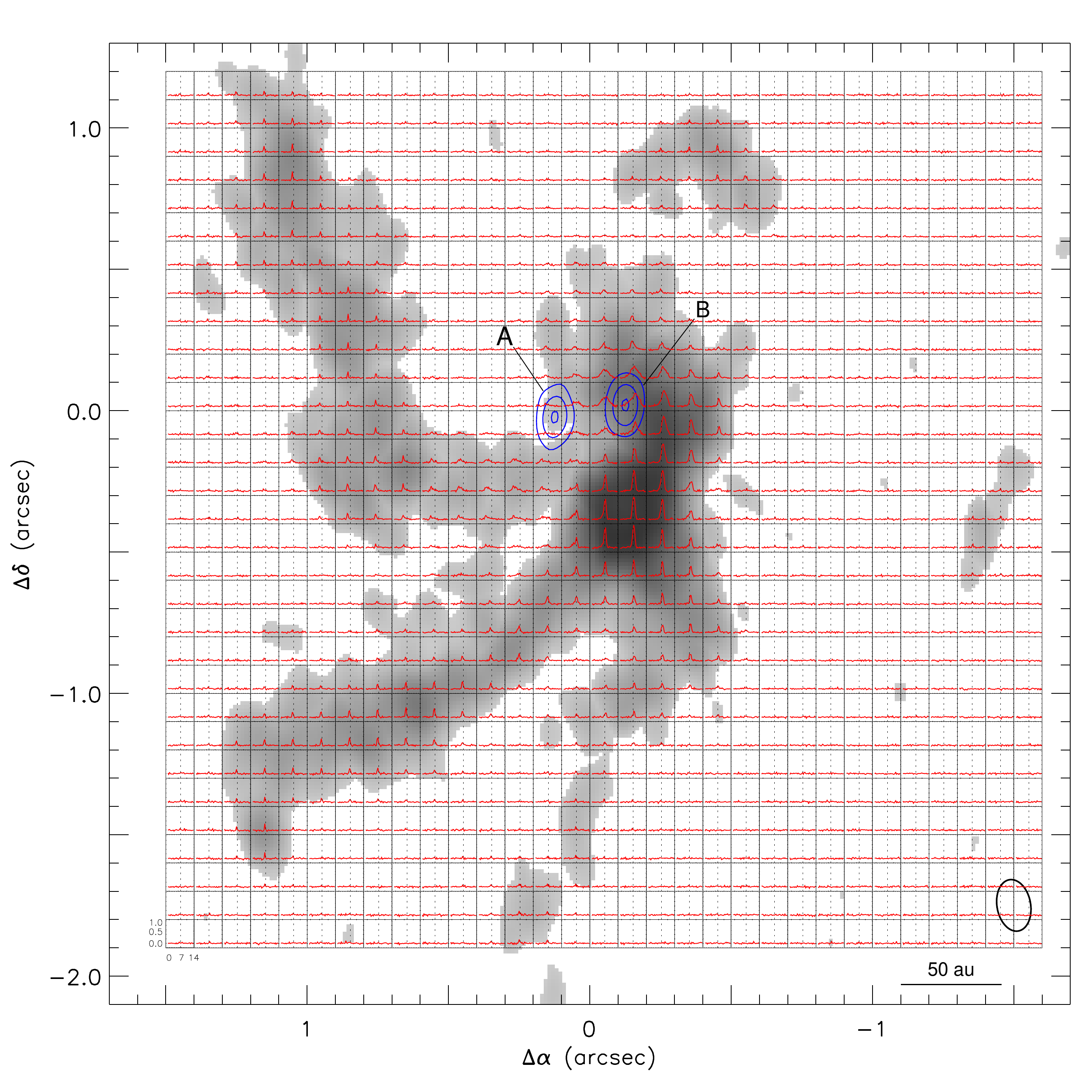}
	\caption{The SO $8_8-7_7$ spectral grid map of IRAS 04239+2436. The background gray image is the SO peak intensity map with the threshold of $6\sigma$ (=0.02 Jy beam$^{-1}$). The blue contours show the two continuum sources, Source A (the eastern source) and Source B (the western source) at 857 $\mu$m. The contours are 6, 18, and 36 mJy beam$^{-1}$.
	The vertical dotted line in each grid indicates the source velocity of 6.5 \kms. The broad SO line profiles are detected around Source B.  The open ellipse in the lower right corner represents the beam size.
}
\label{fig_SO_specmap}
\end{figure*}

\begin{figure*}
\includegraphics[height=10cm]{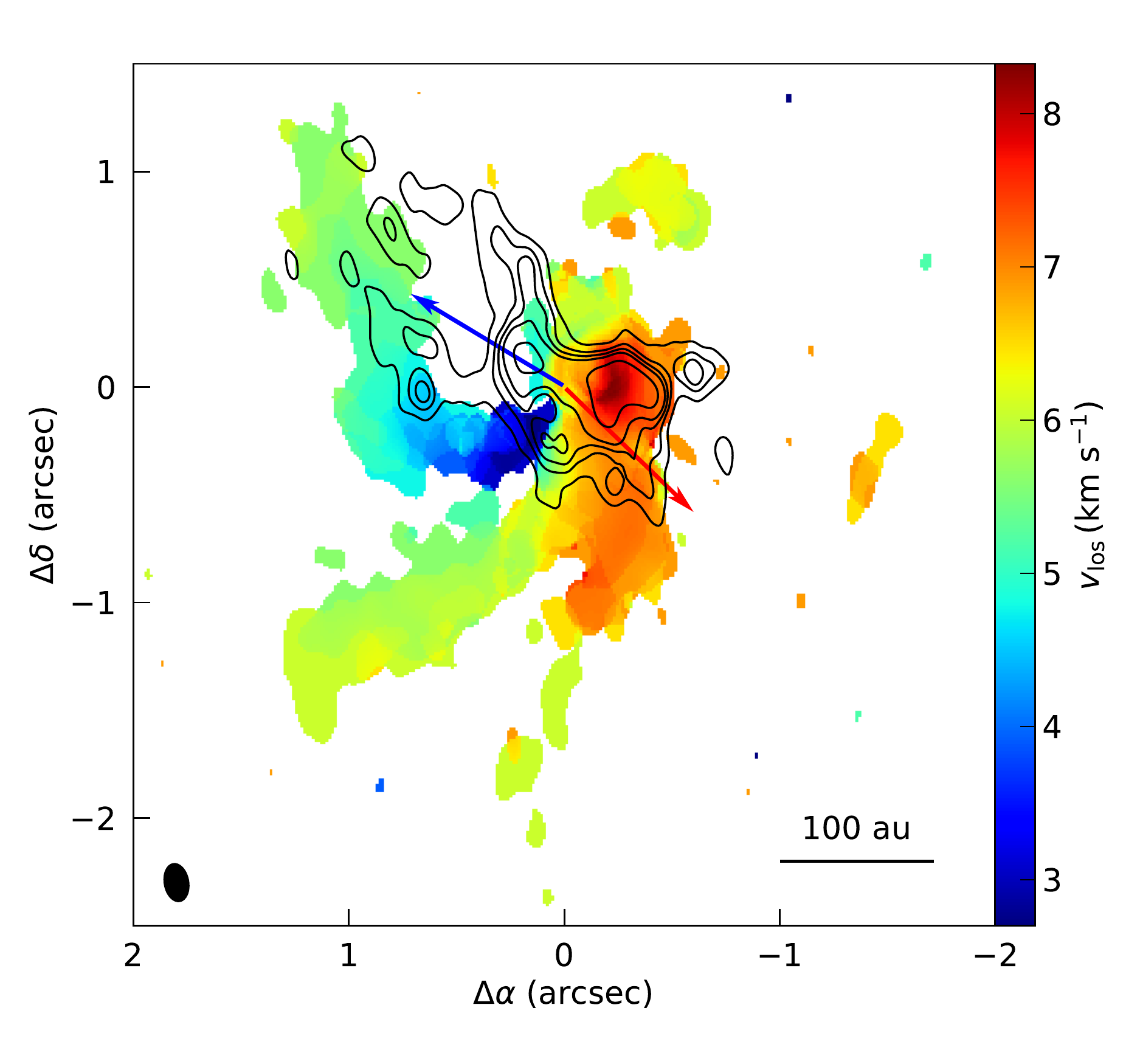}
	\caption{The SO intensity weighted velocity map (colors) and the HCO$^+$ integrated intensity map (contours). The contour levels are 4.0, 5.5, 6.5, 7.5, and 10.0 $\sigma$\ ($\sigma=0.017$ Jy beam$^{-1}$). 
	The SO emission traces the spiral arm structure while the HCO$^+$ emission traces the outflow cavity walls.
	The blue (PA=59$\arcdeg$) and red (PA=226$\arcdeg$) arrows indicate the directions of the [Fe~\textsc{ii}] 1.644~$\mu$m jet of HH 300 \citep{Reipurth2000}.
} 
\label{fig_SO_m1_HCOp_m0}
\end{figure*}

The ALMA observation (Figs.~\ref{fig_SO_mom02}, \ref{fig_HCN_SO}, and \ref{fig_SO_specmap}) of IRAS 04239+2436 reveals two compact continuum sources corresponding to Sources A and B, without any continuum emission associated with the circumbinary material or beyond the circumbinary scale ($\sim$100 au). Unlike the continuum image, the molecular line emission shows 
an extended gas structure up to 400 au; the HCO$^+$ emission distributes mainly along the bipolar outflow cavities (see Fig.~\ref{fig_SO_m1_HCOp_m0}), while the SO line emission distributes along several arm structures (Figs.~\ref{fig_SO_mom02} and \ref{fig_SO_specmap}). 
The most prominent feature is the large extended ($\sim$400 au) multiple (at least three) arm structures traced by the SO line emission. In contrast, the HCN and SO$_2$ emission are 
detected only toward Source B (Fig.~\ref{fig_HCN_SO}), which is known as a secondary in the NIR images.

The large SO molecular spiral arms probably connect the envelope directly to the circumstellar disks around each stellar component.
The arms can be developed by the gravitational torque induced by the dynamical interactions of the binary or multiple protostars \citep{Offner2010,Matsumoto2015}.
This dynamical process generates shocks, and molecular gas can be excited by shock to produce detectable emission along spiral arms.
A recent study for the sulfur chemistry induced by the accretion shocks \citep{vanGelder2021}
demonstrated that even a low-shock velocity ($\sim$3 \kms) in dense ($n \sim 10^7$ cm$^{-3}$) gas can greatly enhance the  abundances of SO and SO$_2$ via the chemical process initiated by the desorption of CH$_4$.
In the physical condition developed by the low velocity shocks, H$_2$S or OCS can be also sublimated from grain surfaces and subsequently react with H, OH, and O$_2$ to form SO \citep{Esplugues2014,Millar1993}.
Therefore, the SO emission detected along the large arm structures in IRAS 04239+2436 is likely originated by shocks generated by gravitational interaction between the infalling envelope and the orbital motion of the triple system (see Section \ref{sec:simulation}). 
In contrast, the large dusty arms are not detected around IRAS 04239+2436 probably because of the low dust column density along the spiral arms at the scales of several hundreds au in the envelope.

\section{Numerical Simulations}
\label{sec:simulation}

\begin{figure*}
\includegraphics[width=\textwidth]{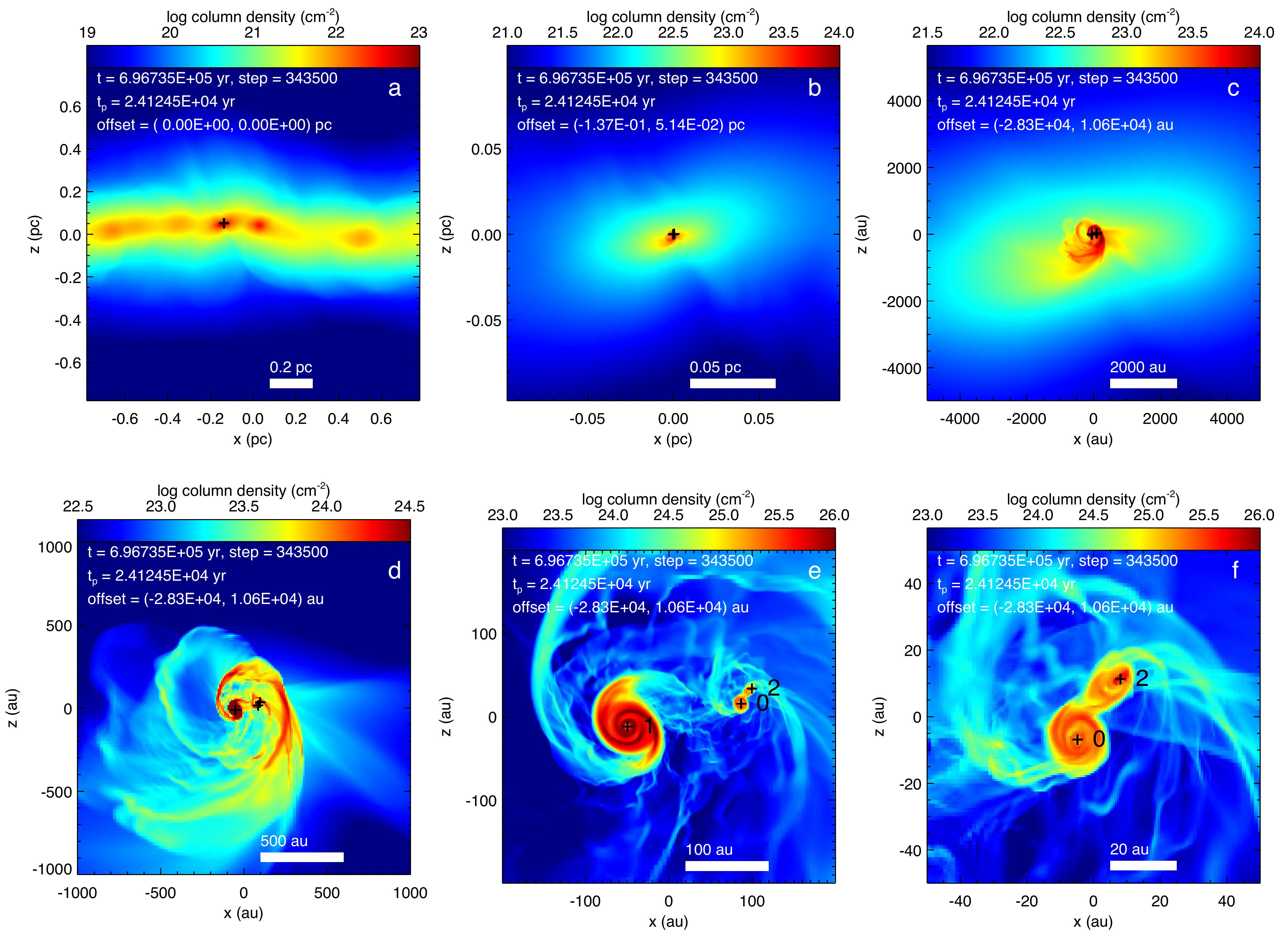}
	\caption{Column density distribution along the $y$-direction on six different scales at the last stage of the simulation ($t = 6.97\times 10^5\,\mathrm{yr}$ and $t_p = 2.41\times 10^4\,\mathrm{yr}$).  The color scales show the column density on a logarithmic scale. The crosses show the positions of the protostars, which are labeled with identification numbers. Column density was obtained by extracting the cubic region from the entire data. Note that each panel shows different size scales. 
} 
\label{fig_colden_zoom}
\end{figure*}

The arm structures confined within dense circumbinary disks toward several binary or multiple systems have been detected in dust continuum \citep{Alves2019,Diaz22}. Theoretical simulations \citep{Bate1997,Matsumoto2019} for a binary system interacting only with a circumbinary disk have shown that two main arms, one of which is connected to each stellar component, are developed, and the secondary component accretes more mass to result in the equal mass binary at the end. However, IRAS 04239+2436 is likely a triple protostellar system that presents very clear triple arms. In fact, the spiral arms developed by triples in a turbulent envelope \citep{Matsumoto2015} have more branches with diverse curvatures, compared to the spiral arms developed inside the circumbinary disk \citep{Matsumoto2019}.
Therefore, in order to explain the observed multiple arm structures in IRAS 04239+2436, we have reanalyzed a hydrodynamic (HD) simulation of a fragmenting turbulent filamentary cloud that forms a multiple stellar system \citep{Matsumoto2015}. 
The adaptive mesh refinement (AMR) technique can zoom into one core where multiple protostars are forming (Fig.~\ref{fig_colden_zoom}).


As an initial condition of the simulation, we consider a filamentary cloud in an equilibrium state, in which the thermal pressure supports the cloud against self-gravity in the computational domain of $(1.56\,\mathrm{pc})^3$.  When we assume the isothermal gas of $T = 10\,\mathrm{K}$ and an infinite cloud length, the density distribution is given by $\rho(R) = \rho_0 [1+ (R/0.05\,\mathrm{pc})^2]^{-2}$ \citep{Stodolkiwicz1963,Ostriker1064}, where $R$ denotes the cylindrical radius.  The scale height of the filament, $0.05 \,\mathrm{pc}$, is chosen based on the observation of the Hershel survey for the filamentary cloud \citep{Arzoumanian2011}. The density on the filamentary axis is given by $\rho_0 = 2c_s^2/(\pi G R_0^2)=1.45\times 10^{-19}\,\mathrm{g}\,\mathrm{cm}^{-3}$ (the corresponding number density is $n_0 = 3.79\times 10^4\mathrm{cm}^{-3}$), where the scale height is $R_0 = 0.05\,\mathrm{pc}$, and the isothermal sound speed is $c_s= 0.19\,\mathrm{km\,s}^{-1}$ for the gas of 10~K.
We impose turbulence with a power spectrum of $P(k) \propto k^{-4}$ as the initial condition, according to the observed size-linewidth relation of $\sigma(\lambda)\propto \lambda^{1/2}$ \citep{Larson1981}, where $k$, $\sigma$,  $\lambda$ are a wave number, linewidth, length scale, respectively.  The average Mach number of turbulence over the computational domain is unity. 

The calculation was performed by the AMR code, SFUMATO \citep{Matsumoto2007}, assuming the barotropic equation of state, $P(\rho) = c_s^2\rho [1 + (\rho/\rho_\mathrm{cr})^{7/5}]$, where the critical density is set at $\rho_\mathrm{cr}=10^{-13}~\mathrm{g\,cm}^{-3}$  \citep{Masunaga1998}.  The self-gravity of gas is taken into account, and the magnetic fields are ignored for the HD simulations. 
Using the AMR technique, the grid is refined in the course of calculation so that the Jeans length of a dense region is resolved by at least 8 cells \citep{Truelove1997}.
In order to mimic a protostar formation, a sink particle is introduced where the density reaches $\rho_\mathrm{sink} = 10^{-11}\,\mathrm{g\,cm}^{-3}$, and several conditions for particle creation are satisfied \citep{Matsumoto2015sink}. Each sink particle accretes the gas around it within the sink radius, $2.45\,\mathrm{au}$, increasing the mass. 

The evolution of the cloud is as follows.  The turbulence disturbs the filamentary cloud, which fragments into cloud cores because of the gravitational instability (Fig.~\ref{fig_colden_zoom}a).  The cloud cores undergo gravitational collapse (Fig.~\ref{fig_colden_zoom}b--c), at the center of which fragmentation produces four protostars (sink particles) in total, and eventually, three protostars survive (Fig.~\ref{fig_colden_zoom}e).  Fig.~\ref{fig_colden_zoom} shows a snapshot at $t_p = 2.41\times 10^4\,\mathrm{yr}$, where $t_p$ is the time elapsed since the first protostar formation.  At this stage, the three protostars exhibit  stable orbits with a close pair with a separation of 20 au (Fig.~\ref{fig_colden_zoom}f) and an outer protostar with a wide separation of $120 - 200 \,\mathrm{au}$.  As shown in Fig.~\ref{fig_colden_zoom}d, the gravitational interaction between the protostars and the infalling envelope gas produces long spiral arms with several hundreds au scale.

\begin{figure*}
\includegraphics[width=\textwidth]{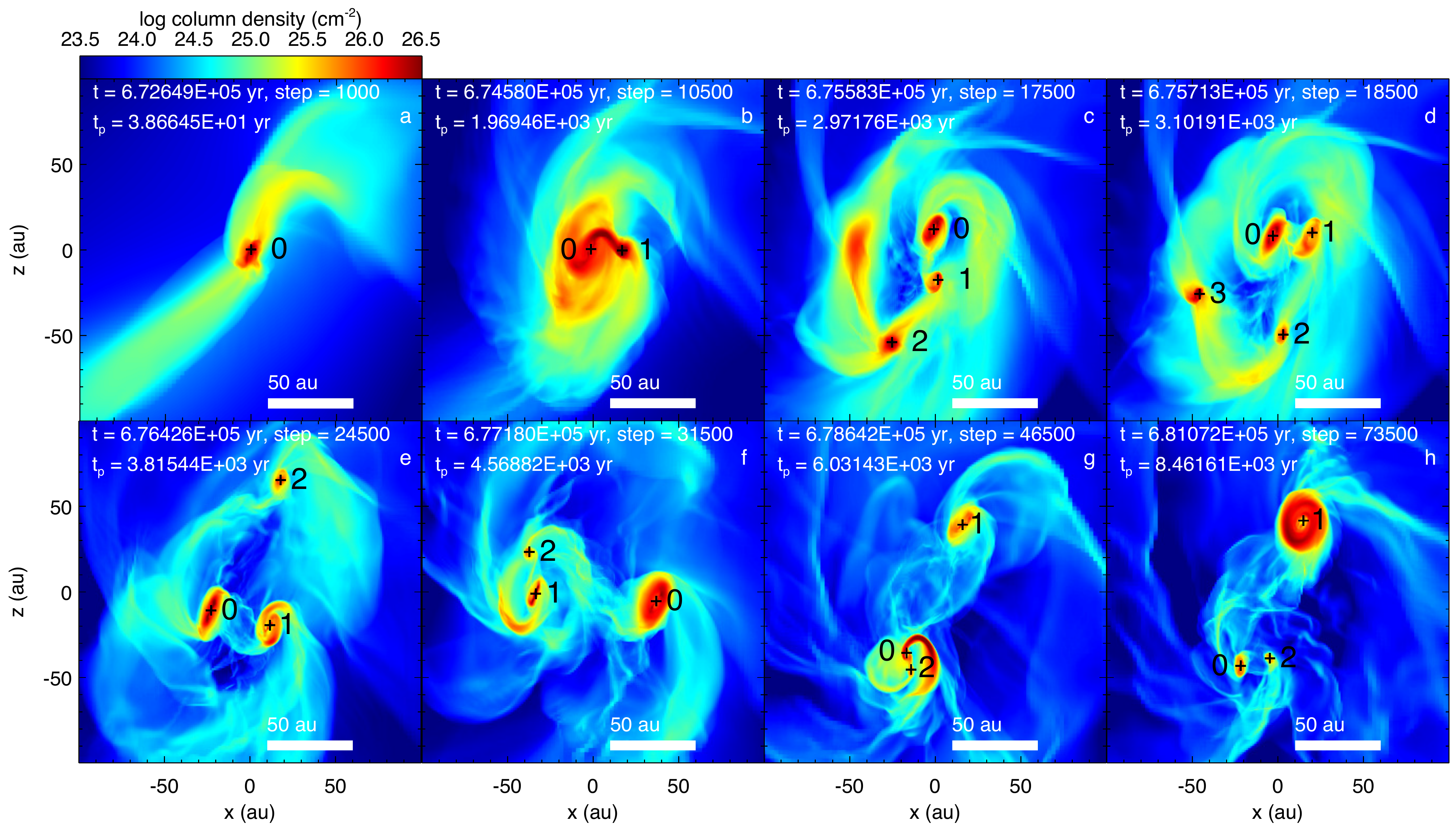}
	\caption{Disk fragmentation at the center of the cloud core.  Eight representative snapshots are shown. Upper panels ({\it a} -- {\it d}) show the formation stages of protostars 1 to 4 from left to right.  Lower panels show the stages of ({\it e}) a merger of protostars 1 and 3, ({\it f}) forming a close binary pair, ({\it g}) exchanging the pair,  and ({\it h}) reaching an almost steady orbit.  The color scale shows the column density distribution on a logarithmic scale, obtained from a box of $(200\,\mathrm{au})^3$.  The protostars are labeled with identification numbers in the order of formation epoch. Each panel is shown centered on the center of mass of the protostars.
} 
\label{fig_fragmentation}
\end{figure*}

In this simulation, all the protostars except for the first one are formed by disk fragmentation. 
Fig.~\ref{fig_fragmentation} shows the time sequence for the formation of protostars and the evolution of their orbits. The first protostar forms in a filamentary structure caused by the turbulence within the cloud core (Fig.~\ref{fig_fragmentation}a).  
The protostar has a circumstellar disk, and the disk fragments to form a second protostar (Fig.~\ref{fig_fragmentation}b).   Then the third and fourth protostars form in succession, one of which merges with the other, and eventually, three protostars survive  (Fig.~\ref{fig_fragmentation}c--e).  The triple stars show chaotic orbits in the early stages, but the orbits gradually stabilize (Fig.~\ref{fig_fragmentation}f--h). The stage of simulation when the overall spiral arm structures are matched well with the observation is $t_p = 1.95\times 10^4\,\mathrm{yr}$.

\begin{figure*}
\includegraphics[width=0.8 \textwidth]{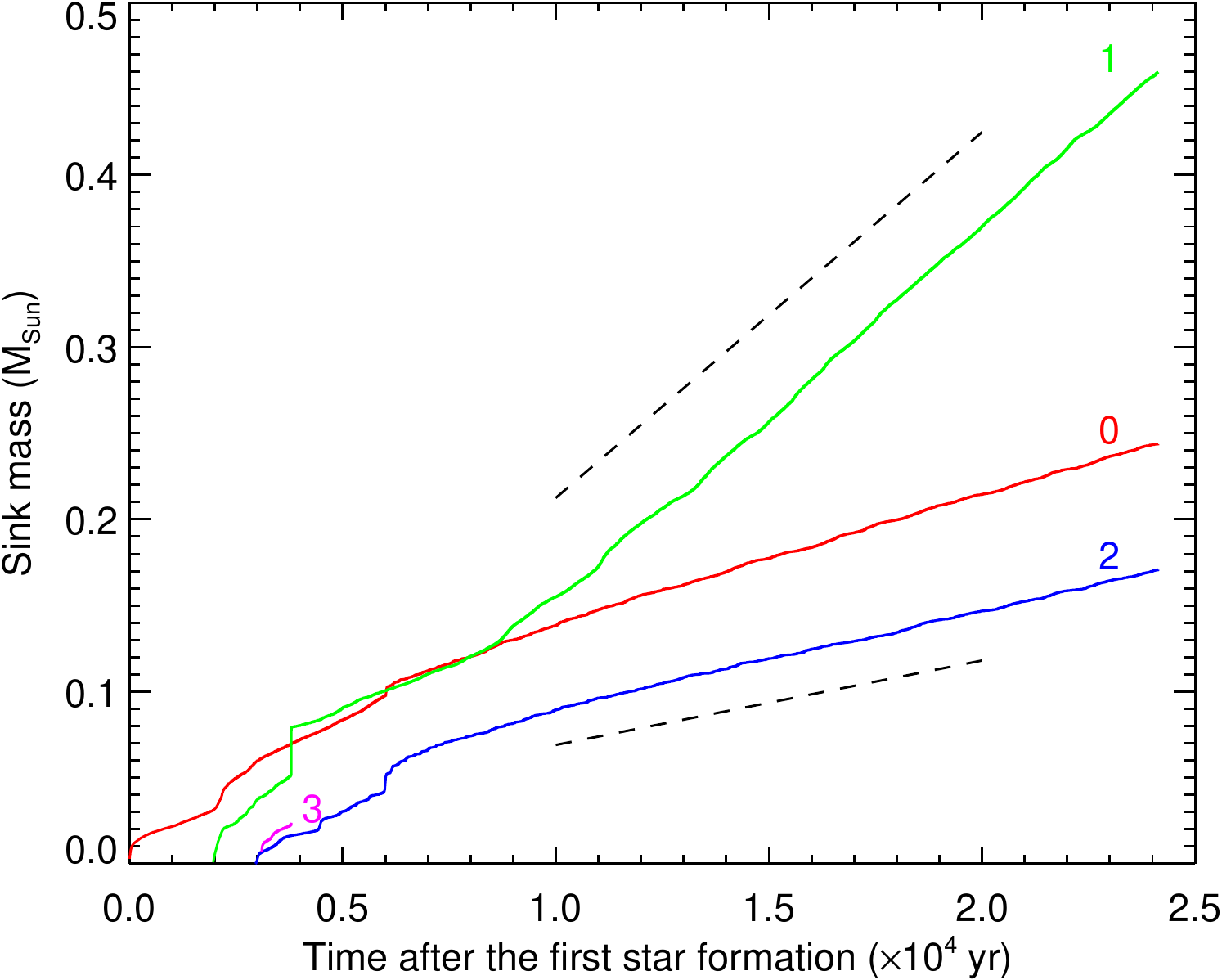}
	\caption{Masses of the protostars as a function of time after the first protostar formation.  The number associated with each line indicates the identification number of each protostar. Dashed lines show relationships of mass accretion rates $\dot{M} = 13 c_s^3/G$ and $3 c_s^3/G$, for comparison. Sink particles 0 and 2 constitute the close binary pairs.
} 
\label{fig_pmass}
\end{figure*}

The protostars are assigned an ID number in the order of formation. Protostar 1 has the most massive disk, with the highest mass and accretion rate (Fig.~\ref{fig_pmass}).  In contrast, protostars 0 and 2 have low accretion rates and smaller masses.  Protostars 0 and 2 form a close binary. Due to a small separation between them, the disks are truncated by dynamical encounters intermittently.  These encounters bring about small disks and low accretion rates, and also misalignment of the disks as shown in Fig.~\ref{fig_snapshot_triple} \citep{Bate2010,Bate2018}, which has been considered as a resulting phenomenon of turbulent fragmentation in the early evolutionary stage of binary formation \citep{Lee2017}. 

\begin{figure*}
\includegraphics[width=0.33\textwidth]{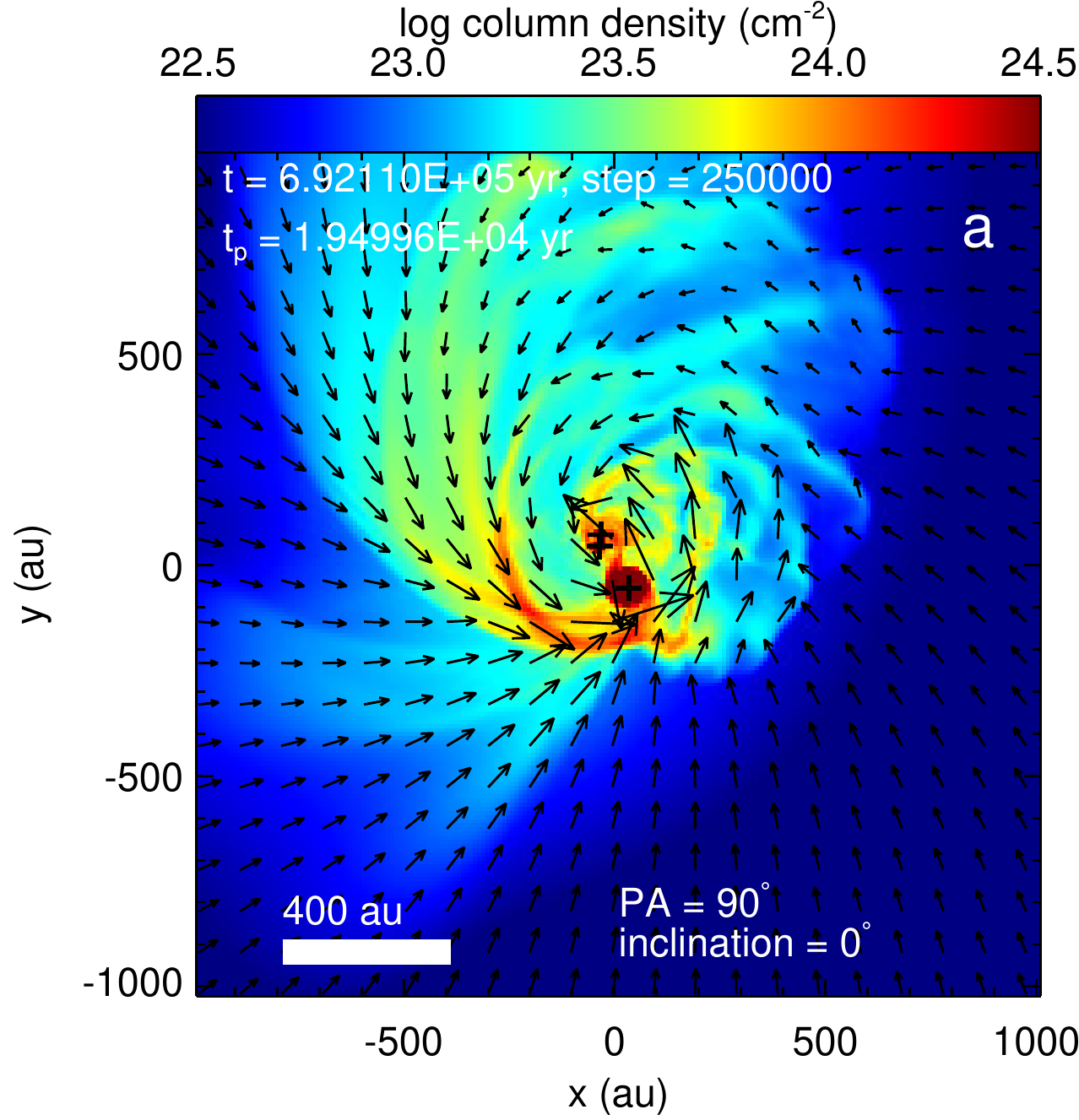}
\includegraphics[width=0.33\textwidth]{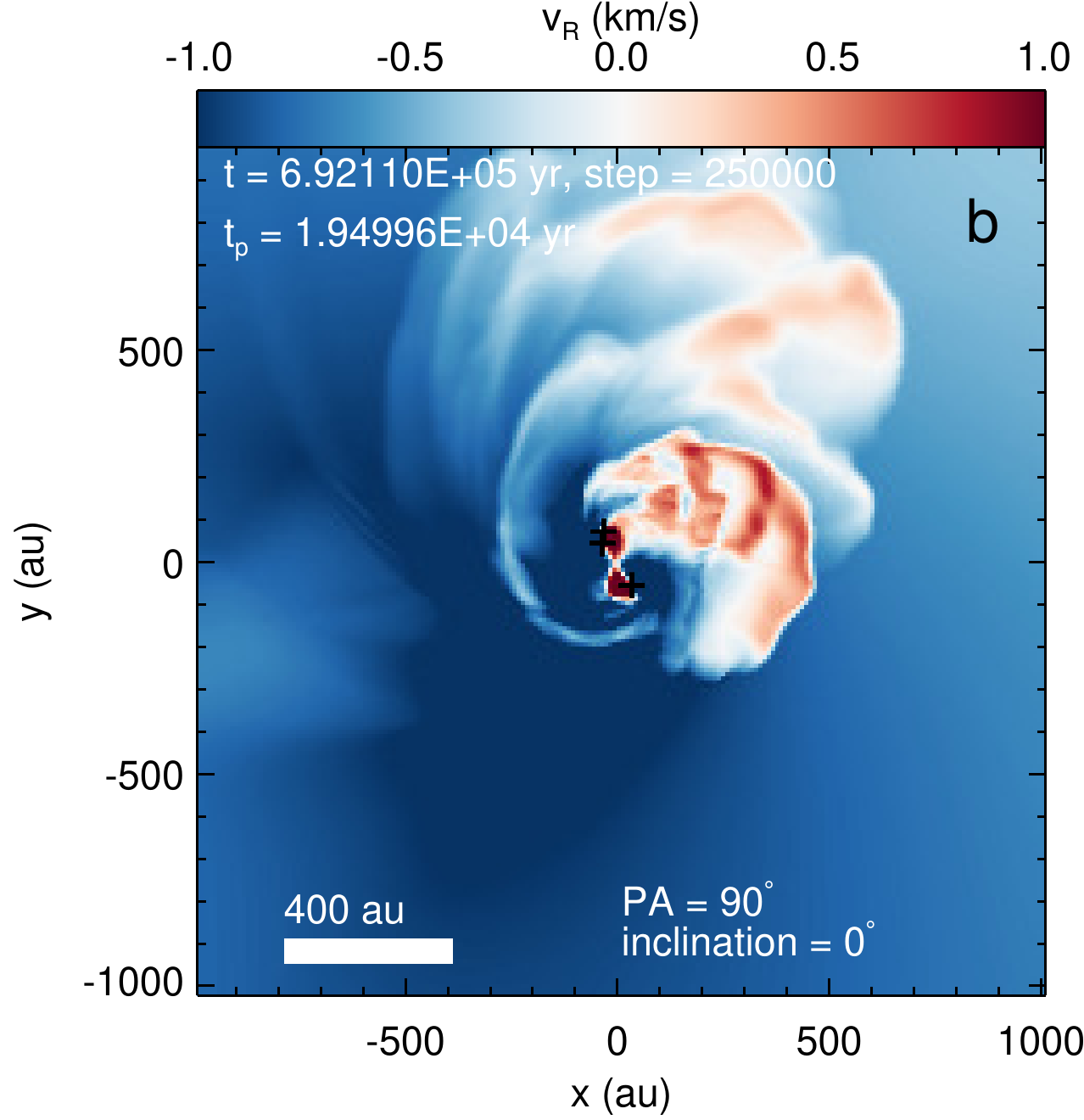}
\includegraphics[width=0.33\textwidth]{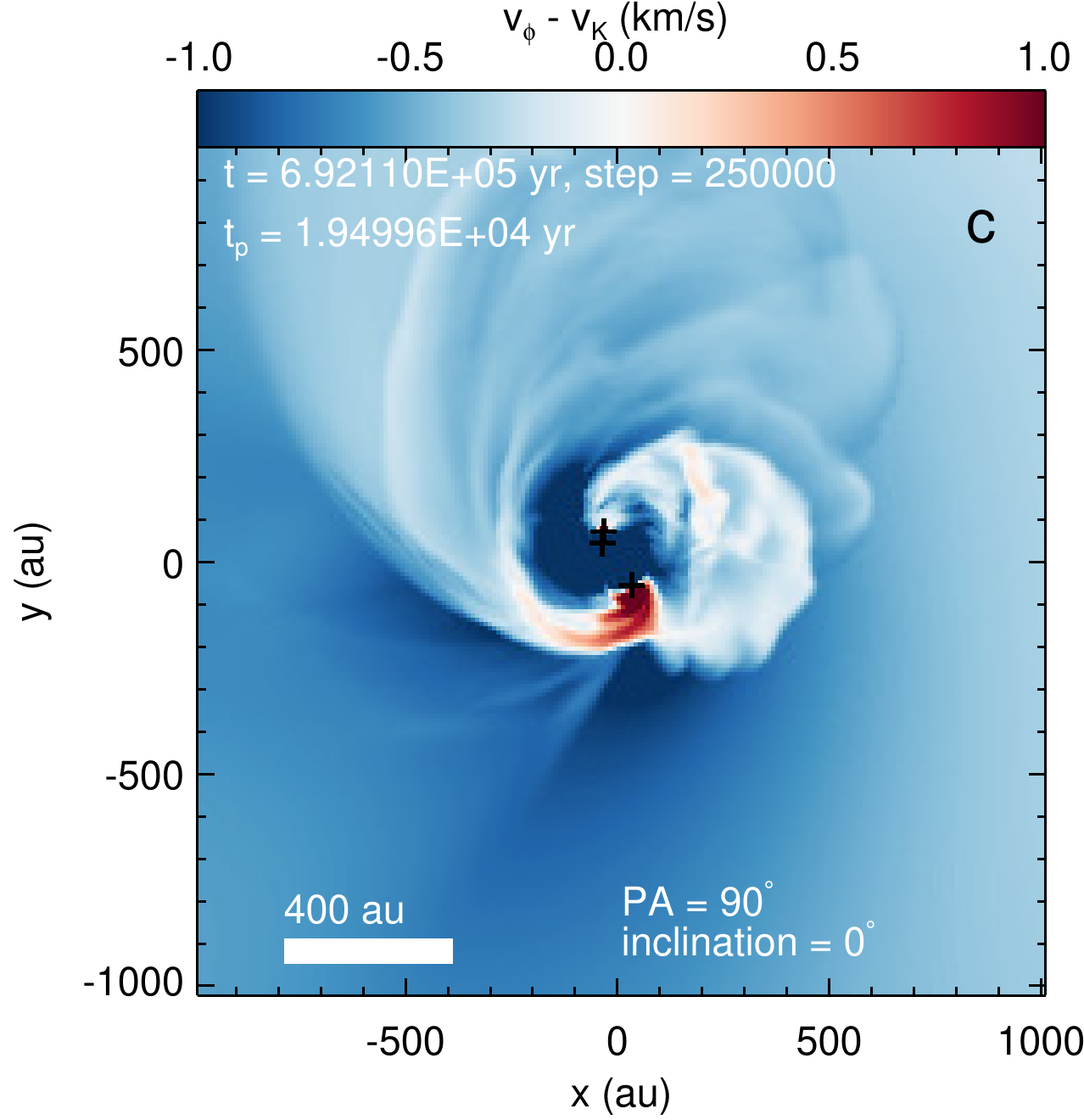}
\caption{
Face on views of the distributions of ({\it a}) column density, ({\it b}) radial velocity, and ({\it c}) rotation velocity excess from the Keplerian rotation, 
at the stage shown in Fig.~\ref{fig_obs_simul}c.  The arrows show the velocity distribution in the left panel.  All the velocity distributions are measured as a relative velocity with respect to the barycenter of the three protostars, and the density-weighted average of the velocity are projected onto the mid-plane in all the panels.
} 
\label{fig_faceon}
\end{figure*}

The interaction between the triplets and the infalling envelope excites arms. The velocity structure of arms is shown in Fig.~\ref{fig_faceon}. In Fig.~\ref{fig_faceon}b, the diffuse arms shown in the upper right of the figure have positive radial velocity (outward velocity), while the prominent arms in the left side have negative radial velocity (infall velocity). The infall velocity along the arms reaches $1-1.5~\mathrm{km\,s}^{-1}$, and the arm structure can be described as infalling streamers, which are narrow structures that asymmetrically feed gas from the envelope to the scale of the circumstellar disks \citep[e.g.,][]{Pineda2020,Bianchi2022,Thieme2022}. The spiral arms also have rotation velocity. The arm extending from the south of the triplets has a high rotation velocity due to the gravity of the protostar (Fig.~\ref{fig_faceon}c). Quantitative velocity analysis along the arm is described in Section~\ref{sec:discussion_comparison}.

\begin{figure*}
\includegraphics[width=0.5 \textwidth]{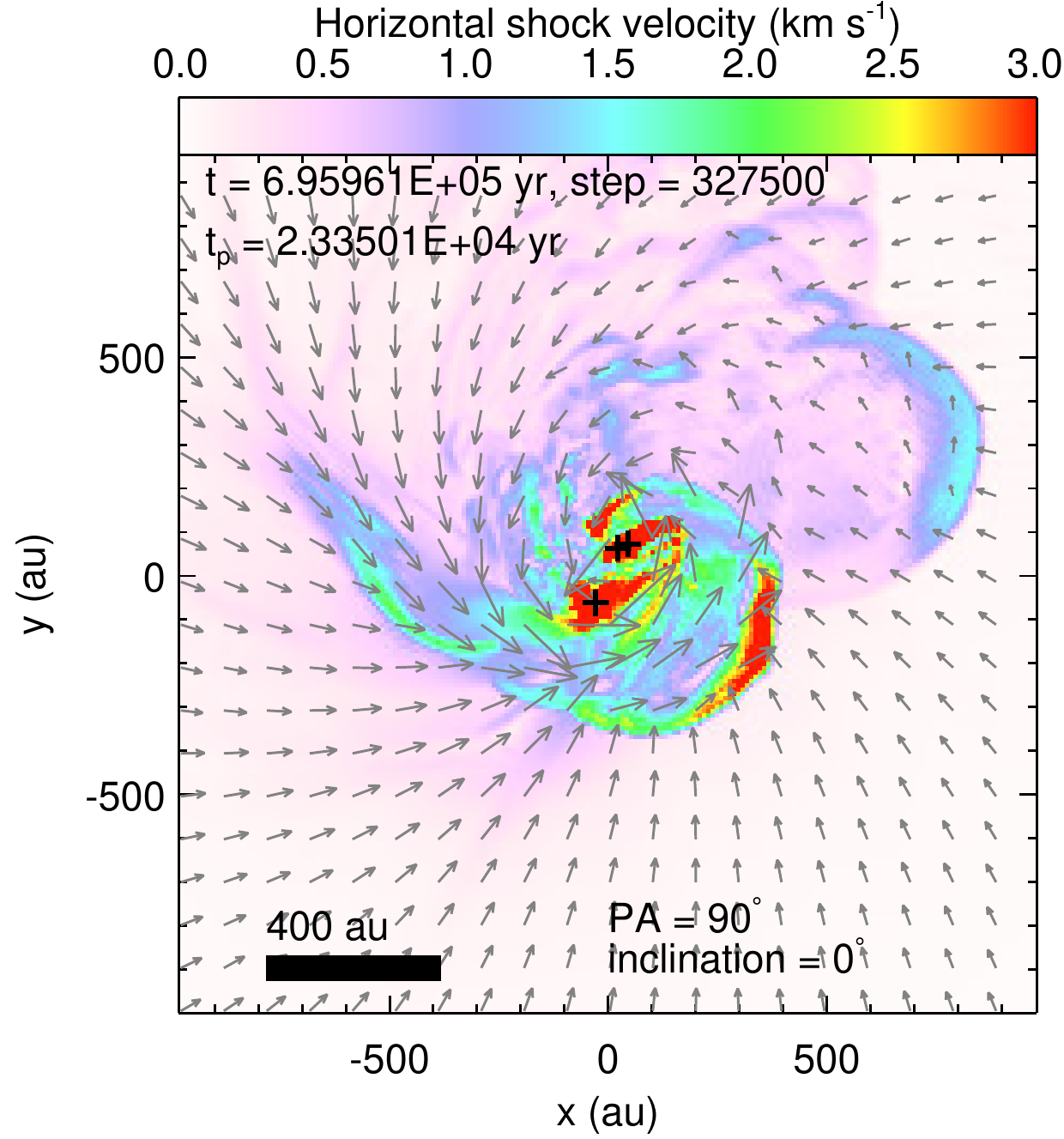}
\includegraphics[width=0.5 \textwidth]{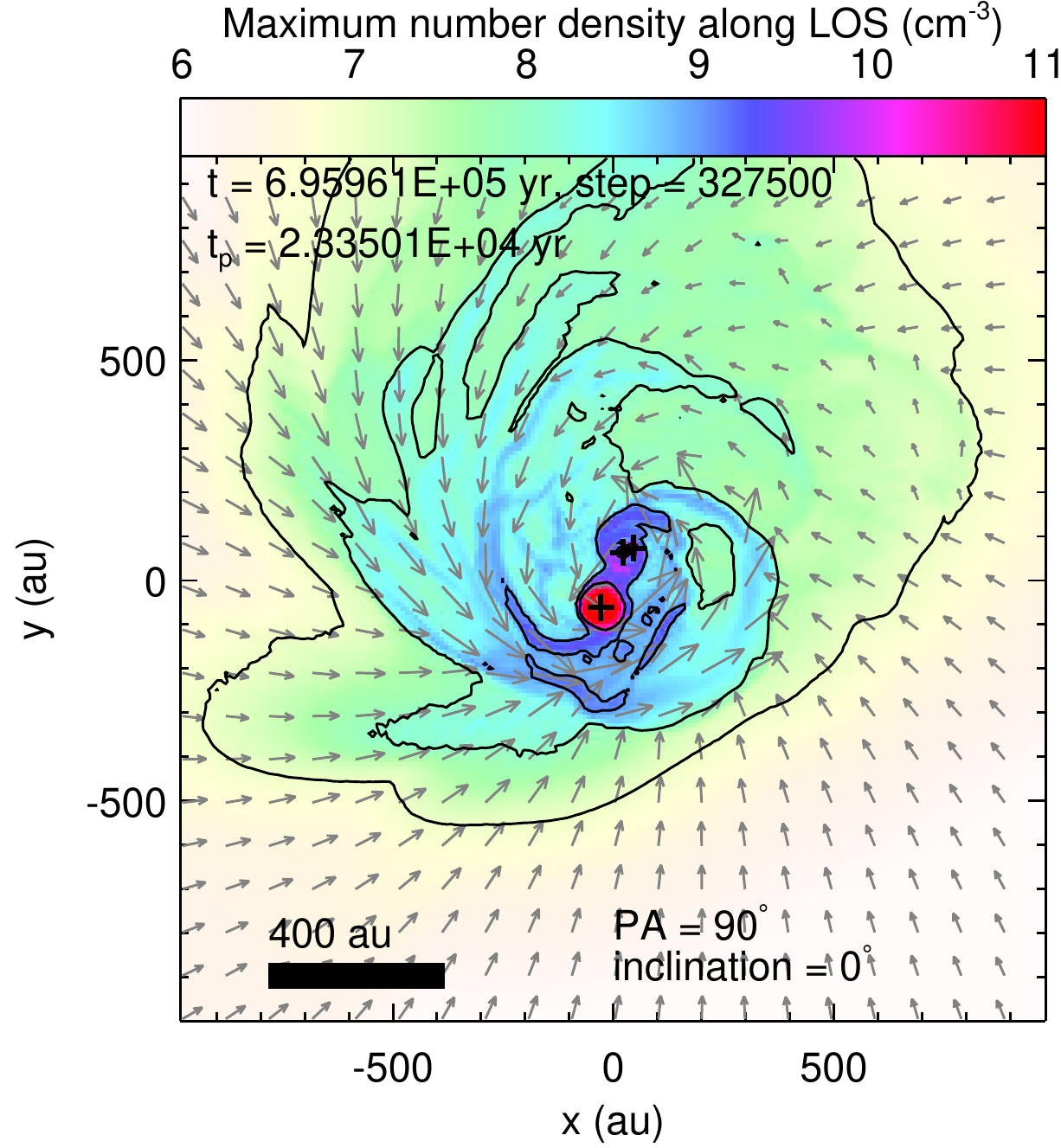}
	\caption{Shock velocity associated with the arms (left) and number density distribution (right). The data is rotated so that it is oriented in a face-on view. The shock velocity is measured as a velocity jump in the horizontal direction. The maximum velocity jump along the z-direction (the line-of-sight) is shown.  The right panel shows the maximum number density along the z-direction.  The contour levels are $n = 10^7$, $10^8$, $10^9$, and $10^{10}\,\mathrm{cm^{-3}}$ in the right panel. The arrows show the density-weighted velocity distribution in both panels. A typical stage that exhibits the shock velocity is shown here and is different from the stage in Fig.~\ref{fig_obs_simul}c.
} 
\label{fig_shockvel}
\end{figure*}

Fig.~\ref{fig_shockvel} shows the shock velocity and density associated with arms.  When the arms are accelerated by the gravitational torque, the shock velocity exceeds $3\,\mathrm{km\,s}^{-1}$ (Fig.~\ref{fig_shockvel}a).  Because an arm accelerated to $\sim 1 \,\mathrm{km\,s}^{-1}$ collides with the infalling gas with $\sim 1 \,\mathrm{km\,s}^{-1}$ in the envelope, the shock velocity may reach $3 \,\mathrm{km\,s}^{-1}$. While the arms propagate into the infalling envelope, the shock velocity decreases to $1.5-2\,\mathrm{km\,s}^{-1}$. The number density in the arms exceeds $10^{7-8}\, \mathrm{cm}^{-3}$ (Fig.~\ref{fig_shockvel}b).  Such velocities and densities of the arms can enhance the SO abundance along the arms \citep{vanGelder2021}.

\section{Discussion}
\label{sec:discussion}

\subsection{A Triple Protostellar System, IRAS 04239+2436}

\begin{figure*}
\includegraphics[width=0.9\textwidth]{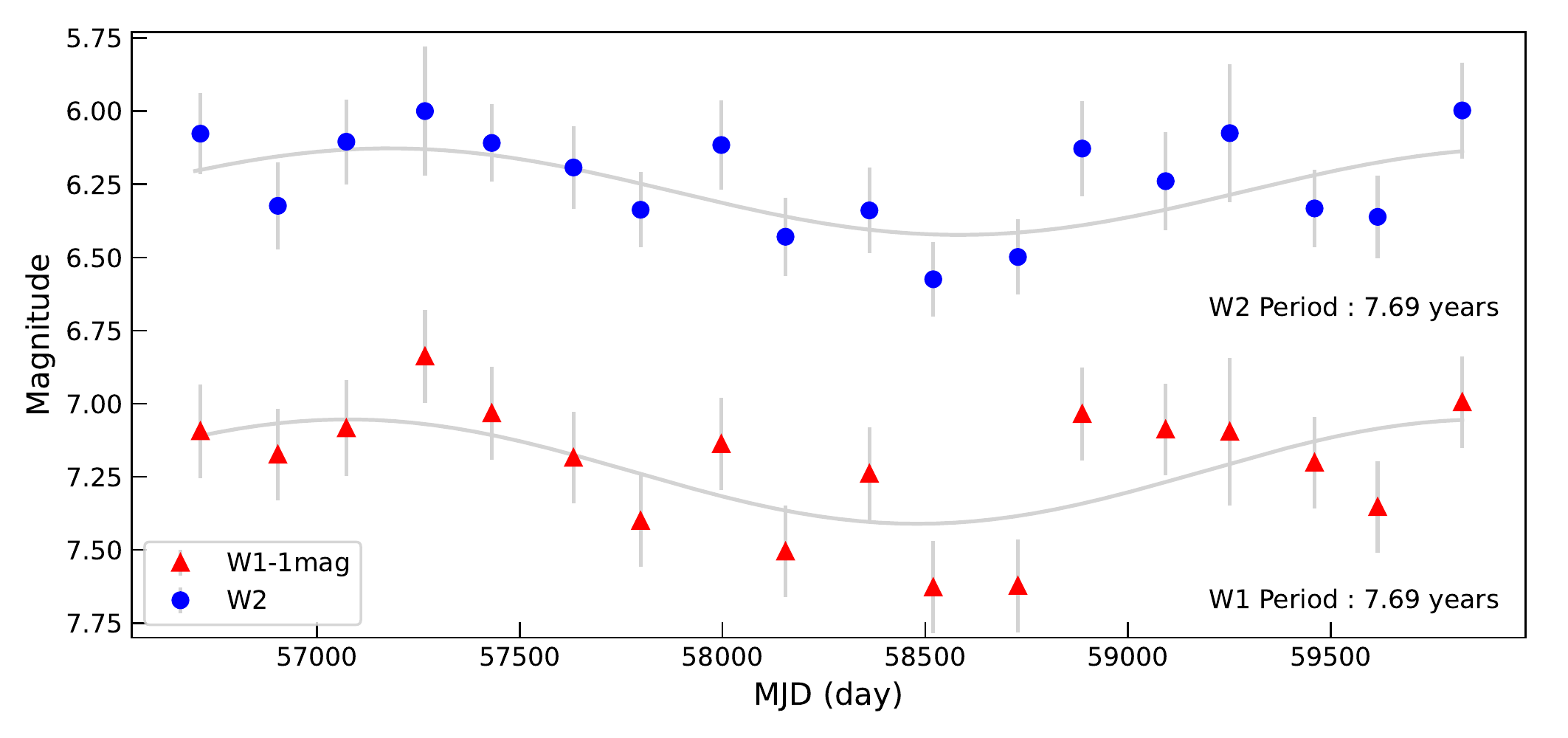}
	\caption{The Near-Earth Object Wide-field Infrared Survey Explorer (NEOWISE) light curves of IRAS 04239+2436 at 3.4 (W1, red) and 4.6 (W2, blue) $\mu$m. We used the NEOWISE-R Single Exposure (L1b) Source Table available at NASA/IPAC Infrared Science Archive \citep{neowise2023}. The periodograms of these light curves present a periodicity of $\sim$8 yrs (gray sinusoidal curves) both in W1 and W2. This period might be associated with the bright [Fe II] jet knots spaced with $\sim$1\arcsec\  \citep{Reipurth2000}. We adopted the method developed by a variability study of young stellar objects \citep{Park21} for the construction of NEOWISE light curves and periodogram analysis, except for including newly released data points for the most recent  four epochs. 
	}
\label{fig_lightcurves}
\end{figure*}

The mid-IR light curve of IRAS 04239+2436 shows a periodicity of $\sim$8 yr (Fig.~\ref{fig_lightcurves}), which is another strong evidence of unresolved close binary interaction within this system, and thus, consisting of triple protostars, indicative of an unresolved very close binary interaction. The anticipated orbital period of the resolved binary is about 270 yr.
The finely collimated strong jet and series of jet knots of Source A in IRAS 04239+2436 invoke the existence of disks fueled from a bigger mass reservoir, i.e., the circumbinary disk or envelope \citep{Jorgensen2022}, despite the small sizes of circumstellar disks truncated by the very close binary interaction within Source A \citep{Reipurth2000,Reipurth2000b}.
In addition, the time intervals between the observed jet knots are 
consistent with the unresolved close binary orbit \citep{Reipurth2000}.

\begin{figure*}
\includegraphics[width=0.5\textwidth]{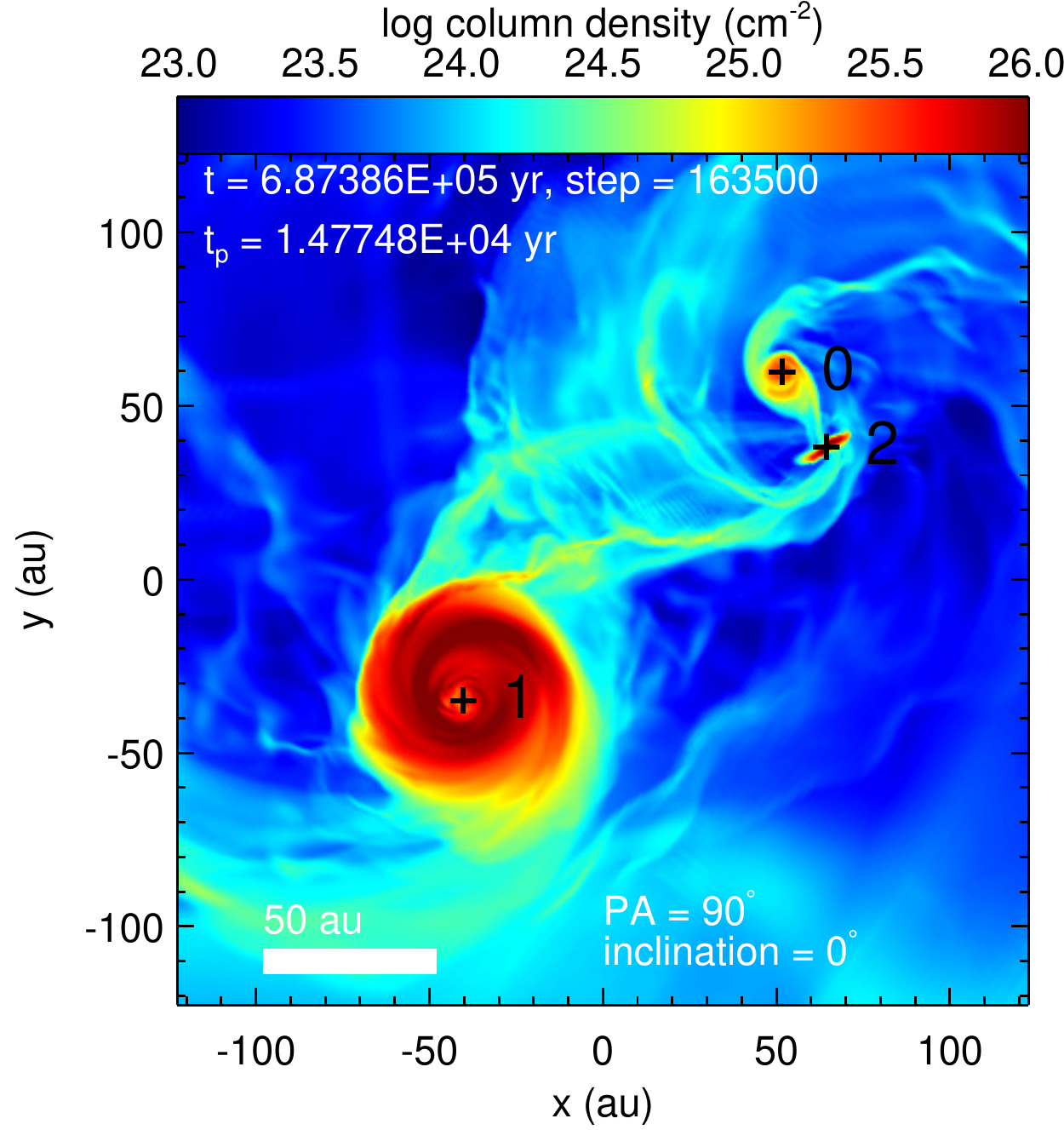}
\includegraphics[width=0.5\textwidth]{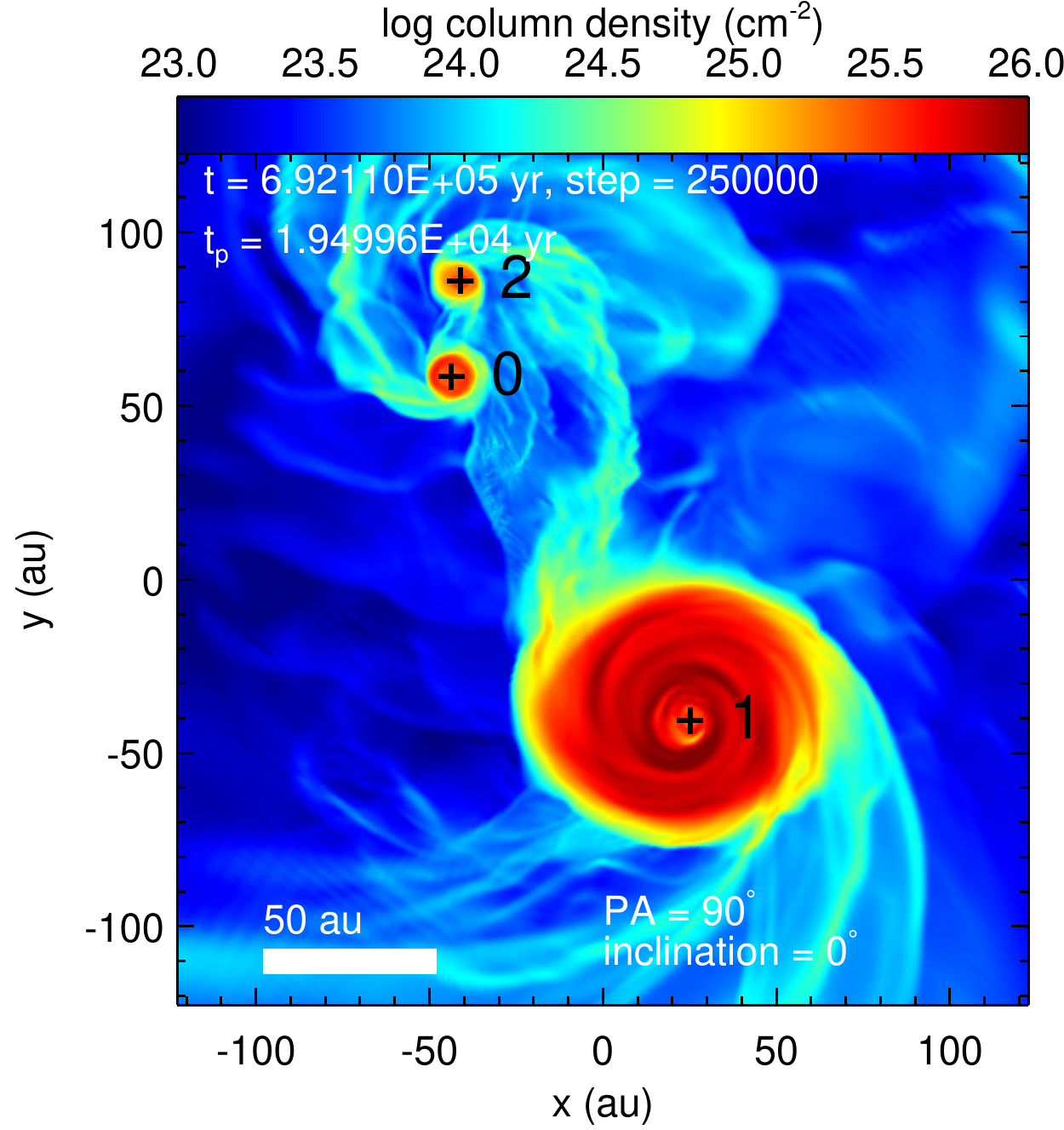}
	\caption{The zoomed-in snapshot images toward the triple system at two time steps. We chose the later time step that describes the best the spiral features of IRAS 04239+2436. At the earlier time step, one of the very close binary shows a misaligned disk rotation axis while the misalignment is reduced at the later time step. The mass of the massive disk is 20\% of the central stellar mass, while the less mass disks have their masses as low as 1\% of the central stellar masses.} 
\label{fig_snapshot_triple}
\end{figure*}

In the NIR images, Source A, which is likely an unresolved very close binary as mentioned above, is brighter than Source B.
However, the comparison with the numerical simulation suggests that Source B is more massive component (Figs.~\ref{fig_pmass} and \ref{fig_obs_simul}) with a larger disk (Fig.~\ref{fig_snapshot_triple}), which can obscure the NIR emission effectively. 
Another triple system with spirals, L1448 IRS3B \citep{Tobin2016} also shows that the dimmer component is more massive; unlike IRAS 04239+2436, the more massive one is the close binary in L1448 IRS3B.
The HCN line emission is detected only toward Source B, consistently supporting the existence of a very massive disk around Source B (Fig.~\ref{fig_HCN_SO}b), which is likely actively accreting gas along the spiral arm 2 (Figs.~\ref{fig_SO_specmap} and \ref{fig_obs_simul}a).  According to the simulation (see Fig.~\ref{fig_snapshot_triple}), the column density of the larger disk is higher, resulting in its mass being higher than the other two close disks by a factor of 100. In addition, the accretion rate in the primary is higher than those of the secondary and tertiary by more than a factor of three. This higher accretion rate to the more massive protostar heats its disk more efficiently. As a result, the HCN line intensity, which is proportional to the column density and temperature, is strong enough to be detected in Source B.

This accretion streamer associated with Source B and arm 2 is traced by another shock tracer, the SO$_2$ emission (Fig.~\ref{fig_HCN_SO}c--d) as well as the broad SO line profiles around Source B (see Fig.~\ref{fig_SO_mom02}b and the broadened line profiles in Fig.~\ref{fig_SO_specmap}). 
In contrast, the 857 $\mu$m continuum fluxes toward Sources A ($51.5 \pm 0.3$~mJy) and B ($52.8 \pm 0.3$~mJy) are comparable. 
This could be due to the continuum optical depth.
In the massive disk around Source B, grains easily grow to millimeter sizes and become optically thick at millimeter wavelengths \citep{Harsono2018,Lee2019}. As a result, the submillimeter emission may not trace the total disk mass, as shown in the ALMA observation of L1448 IRS3B \citep{Tobin2016}.

\subsection{Comparison between Observation and Simulation}
\label{sec:discussion_comparison}
\begin{figure*}
\includegraphics[width=1\textwidth]{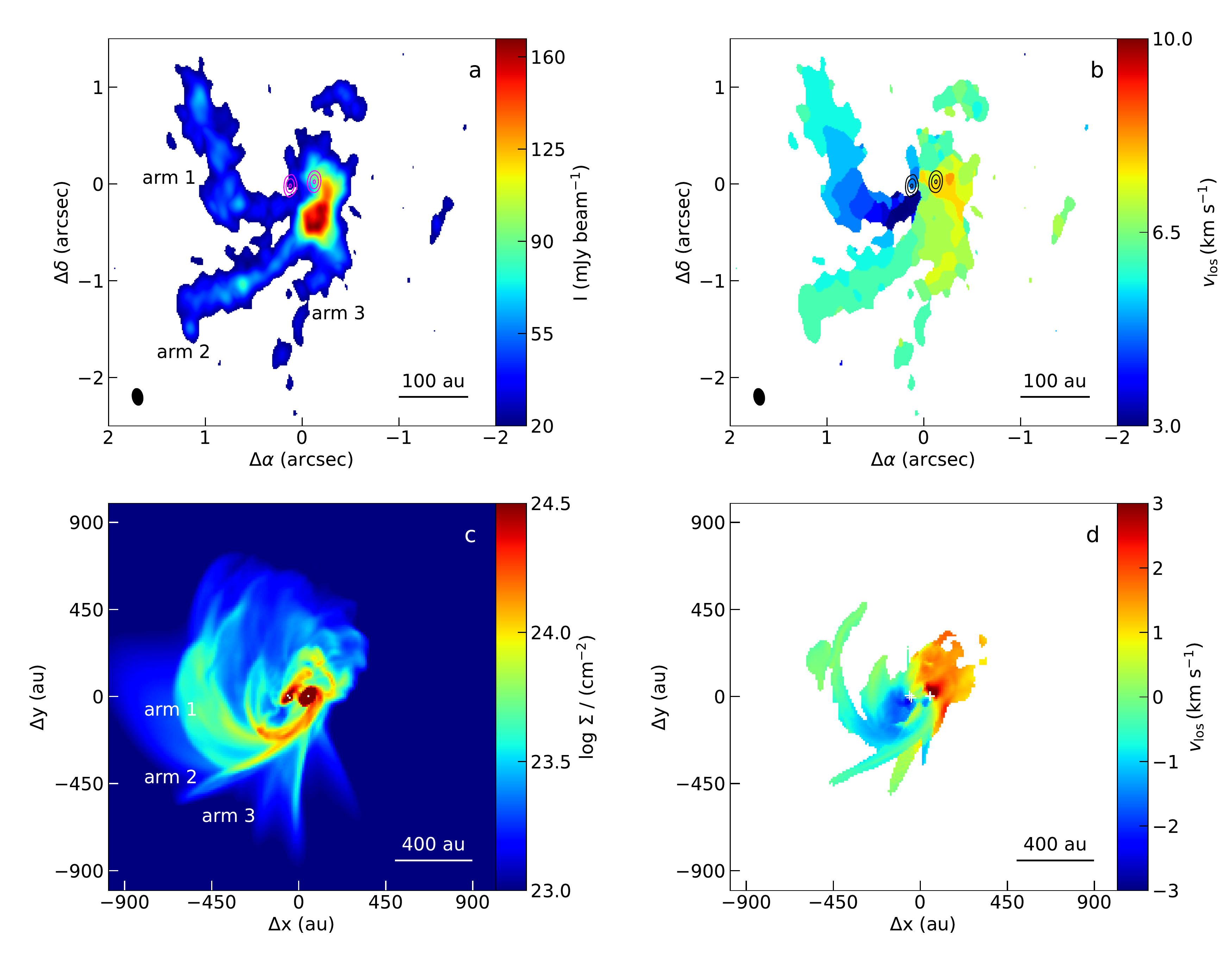}
	\caption{Comparisons between observation and simulation. (a) The SO $8_8$--$7_7$ peak intensities in units of \mjybeam\ and (b) the velocities of the intensity peaks in units of \kms\ toward IRAS 04239+2436. The source velocity is $V_{\rm LSR} = 6.5$~\kms.
	The SO peak intensity and velocity maps were generated using the threshold of $6\sigma$ (0.02 Jy beam$^{-1}$). The contours in (a) and (b) present the continuum emission at 857 $\mu$m. (c) The gas column density distribution and (d) the velocity distribution of the numerical hydrodynamic simulation. 
	The arms clearly identifiable are named in (a), and the arms that may correspond to the simulation are marked in (c). The white dots (c) and crosses (d) mark the positions of the three protostars. The simulation was rotated and projected consistently with IRAS 04239+2436.
}
\label{fig_obs_simul}
\end{figure*}

\begin{figure*}
\includegraphics[width=0.5\textwidth]{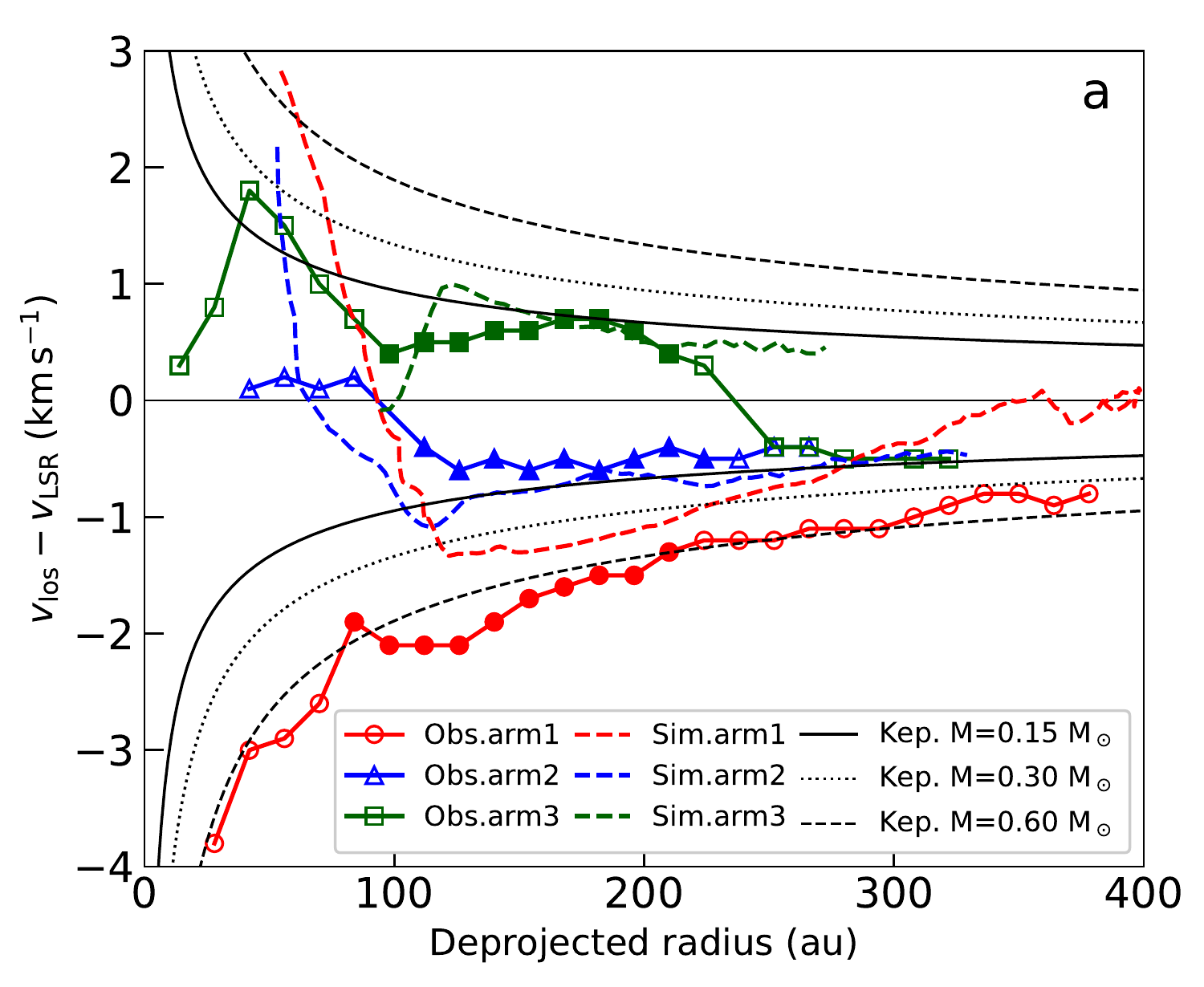}
\includegraphics[width=0.5\textwidth]{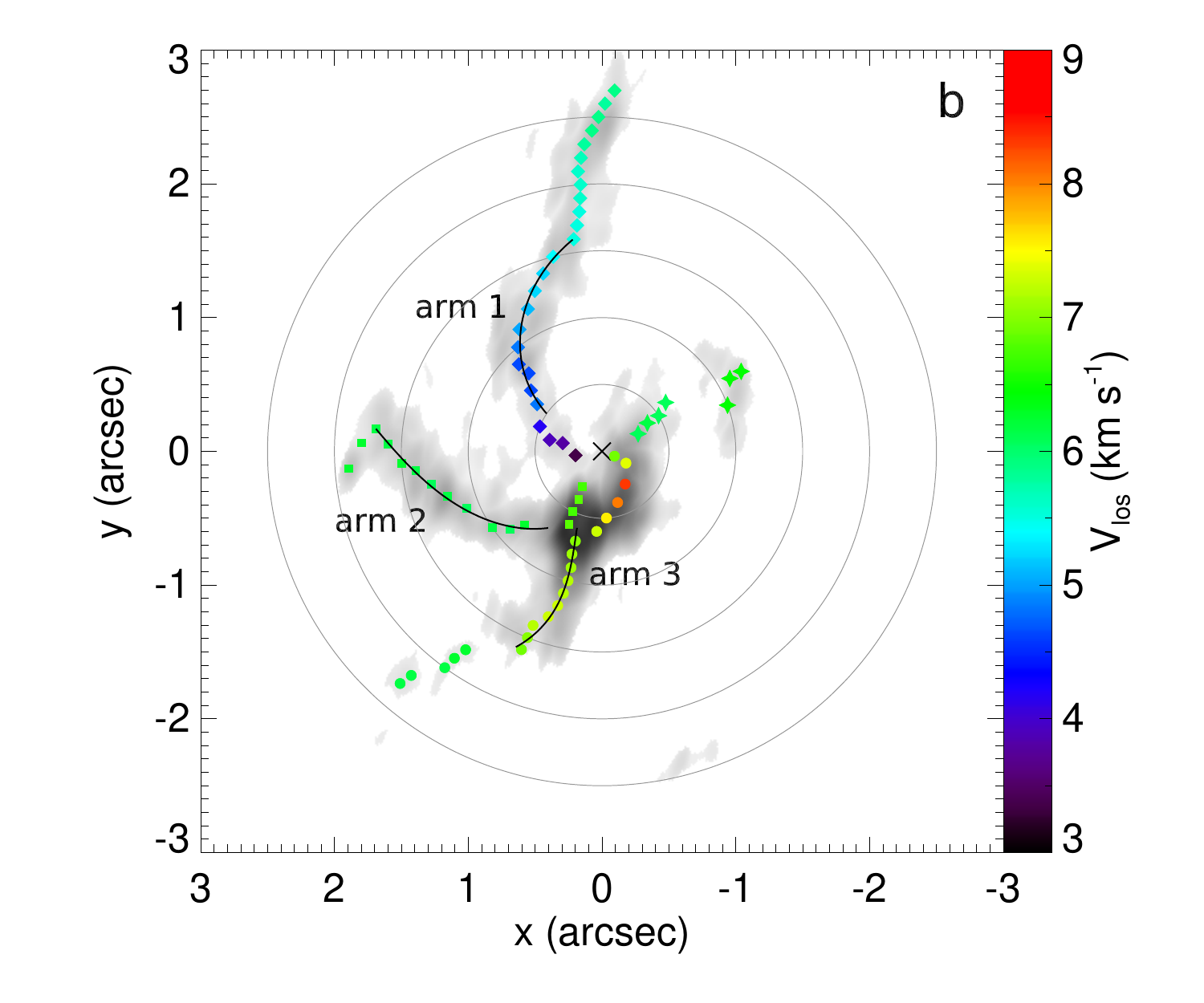}\\
	\caption{Kinematics of the spiral arms of IRAS 04239+2436. (a) Peak velocities (relative to source velocity) along the spiral arms. 
	Color solid lines with symbols are the velocity at the peak-intensity positions of the spiral arm features (Fig.~\ref{fig_obs_simul}) of IRAS 04239+2436. 
    Filled symbols are the velocities at the positions used for the polynomial fittings as presented with solid black curves in (b).
	Color dashed lines are the velocity of 
	the arm features from the simulation of which deprojected radii are scaled by a half to match the total mass of companions. 
	Black lines are the Keplerian rotation curves of 0.15~\msun\ (solid), 0.30~\msun\ (dotted), and 0.60~\msun\ (dashed).
    (b) Peak-intensity positions and velocities of the spiral arms (colors) on the deprojected peak intensity map of SO $8_8-7_7$ (gray image). 
	Symbol `$\times$' marks the center position defined by the center of mass of the two continuum sources assuming an equal mass. 
	The black solid lines indicate the polynomial fitting results over the peak intensity positions that may belong to the arm structures with coherent kinematics. 
	Gray circles are drawn from 0.5$\arcsec$ to 2.5$\arcsec$ in steps of 0.5$\arcsec$.
}
\label{fig_kinematic}
\end{figure*}

While the total protostellar mass and their separation are greater than those observed toward IRAS 04239+2436 \citep{Lee2016}, we emphasize that the overall arm features (Fig.~\ref{fig_obs_simul}) and their kinematics (Fig.~\ref{fig_kinematic})  
produced by the simulation are very similar to the observed ones. Fig.~\ref{fig_kinematic} compares the kinematics between observation (solid lines) and simulation (dashed lines) along the spiral arms to show that the simulated dynamical process governs the observed arm structures.
In Fig.~\ref{fig_obs_simul}, the total gas distribution is presented for the simulation, while the observed arm structures in IRAS 04239+2436 are revealed by the SO emission, which traces the shocked gas. 
The peak intensities of the continuum in the circumstellar disks are about 40 mJy beam$^{-1}$, and the rms noise level is about 0.1 mJy beam$^{-1}$. In order to detect the spiral arms in the dust continuum, the column density of spiral arms must be greater than 1\% of the peak column density of the circumstellar disks. While the column densities of simulated spiral arms are lower than that, they satisfy well the physical conditions for the SO chemistry \citep{vanGelder2021} as seen in Fig.~\ref{fig_shockvel}. 

Fig.~\ref{fig_kinematic}a presents the velocity at the peak-intensity position (hereafter, peak velocity) along each spiral arm feature in the SO emission as a function of deprojected radius (symbols connected by solid color lines). The source velocity is $V_{\rm LSR} = 6.5$~\kms\ adopted from C$^{18}$O 1--0 and C$^{17}$O 1--0 observations \citep{Fuller2002}.
The peak-intensity positions were measured on the peak-intensity map of SO $8_8$--$7_7$, which is deprojected by an inclination angle of 55$\arcdeg$ and PA of 140$\arcdeg$ (Fig.~\ref{fig_kinematic}b). 
Because a circumbinary disk has not been detected,  
we assume
55$\arcdeg$ and 140$\arcdeg$ based on the inclination  (50$\arcdeg$--61$\arcdeg$) and PA (137$\arcdeg$--145$\arcdeg$) measured from the continuum sources. 
We also assume that all the spiral arms are on the same plane.
For a deprojected radius from 0.1$\arcsec$ to 2.7$\arcsec$ in steps of 0.1$''$, 
an intensity profile along the azimuthal direction was extracted, and 
peak-intensity position was measured. At each deprojected radius,
the line-of-sight velocity ($V_{\rm los}$) was derived from the SO $8_8$--$7_7$ data cube.
In Fig.~\ref{fig_kinematic}b, the peak-intensity positions of each arm are marked 
by symbols in different colors that represent their velocities as given by a color bar. 
The center position is the center of mass of Sources A and B on the assumption of equal mass. 
As seen in the figure, the peak positions mostly well trace the spiral arm features, but arms 2 and 3 are not clearly distinguishable in the inner region 
($r \lesssim\ 0.7\arcsec$ or $r \lesssim\ 100$ au) where SO emission is the brightest, and the two arms are likely overlapped or colliding. Considering the complex emission around 
the center and the angular resolution ($\sim 0.1\arcsec$) of the observation, the peak velocity at $r < 70$ au may not be reliable.

Fig.~\ref{fig_kinematic}a also presents the velocities of the arms derived from our simulation (dashed color lines) and the Keplerian rotation (black lines) for comparison. Since the total mass and separations among companions 
from the simulation are larger than those of IRAS 04239+2436 by a factor of two, the deprojected radii of the simulation are scaled by a half in the figure.
As the overall arm features of IRAS 04239+2436 and the simulation are similar (Fig.~\ref{fig_obs_simul}), their velocity profiles also show similar trends (Fig.~\ref{fig_kinematic}a). 
The velocity of the arms is not explained by a single Keplerian rotation curve, although arms 1 and 2, beyond 100 au, roughly trace the Keplerian rotation caused by the mass between 0.6~\msun\ and 0.15~\msun. 
In addition, the observed structure of arm 1 beyond 1.5\arcsec\ is not fitted by a polynomial function simultaneously with the inner part of the arm as presented in Fig.~\ref{fig_kinematic}b. 

\begin{figure*}
\includegraphics[width=0.51\textwidth]{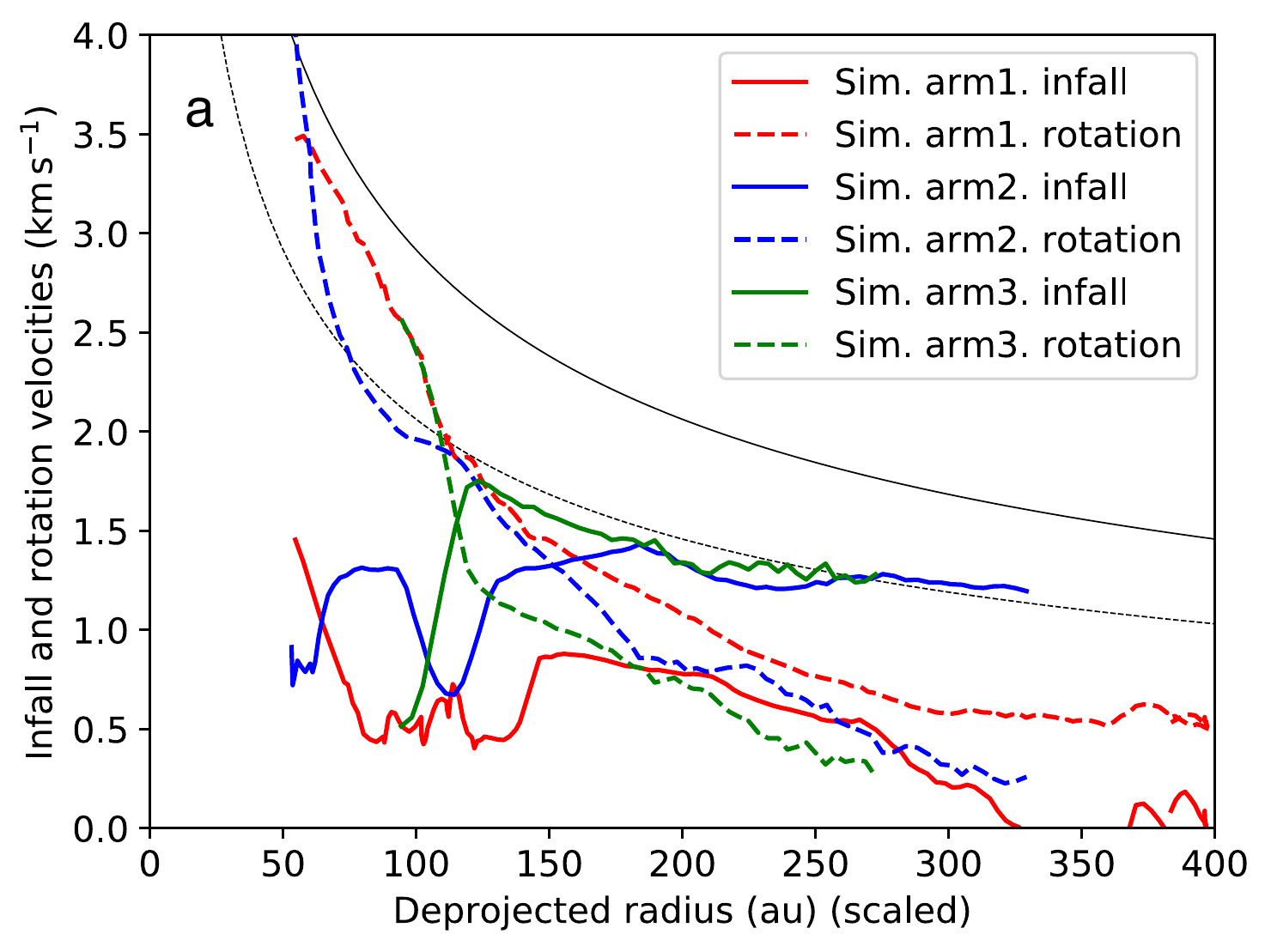}
\includegraphics[width=0.47\textwidth]{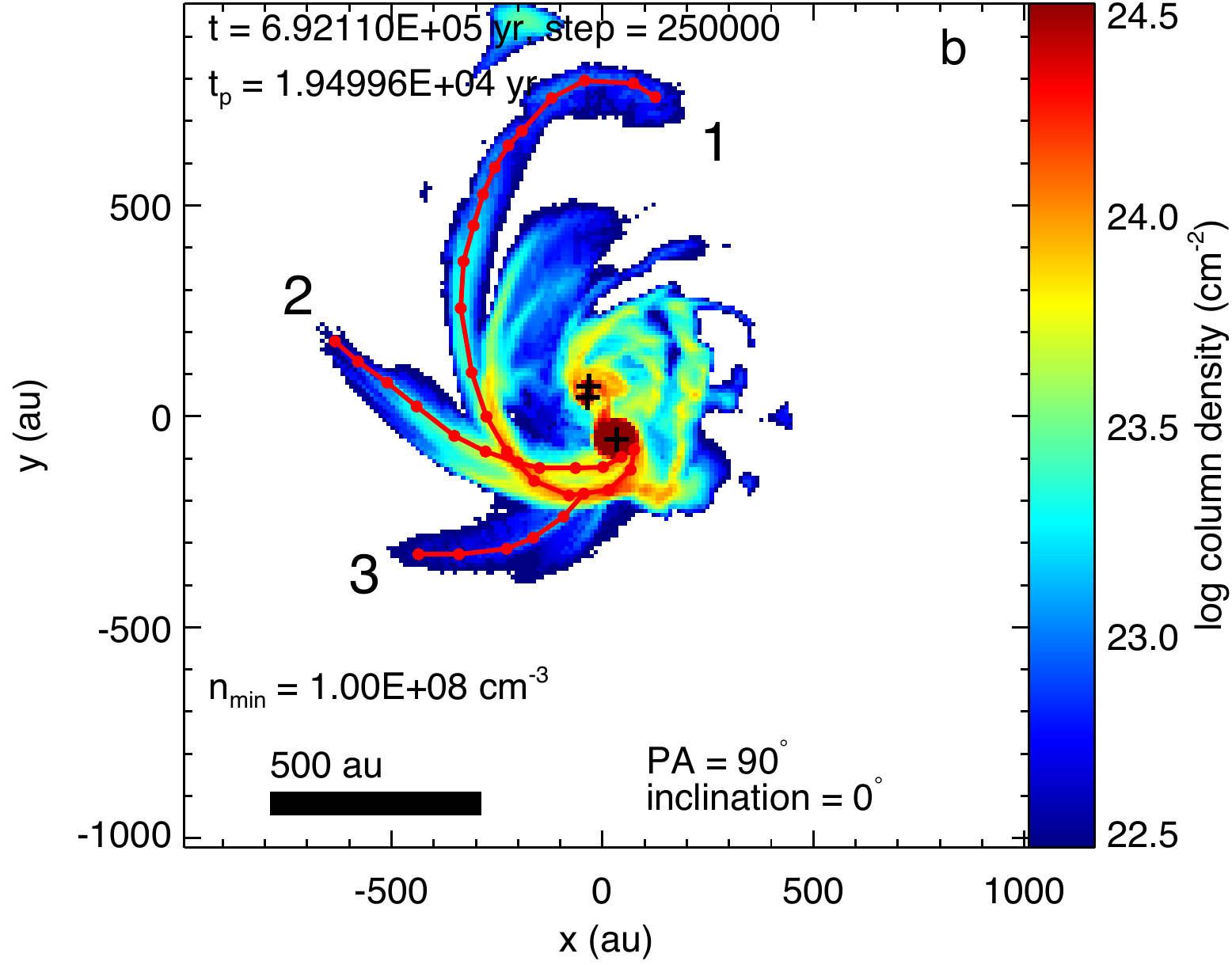}
	\caption{
	(a) Infall and rotation velocity distribution along the arms as a function of radius for the simulation. For comparison, black lines are shown for the freefall velocity (solid) and the Keplerian rotation velocity (dashed) for the total mass of sink particles in the simulation ($0.71 \msun$). The radius is scaled to fit Figure~\ref{fig_kinematic}b. (b) Configuration of the spiral arms (red curves) for the simulation, which is shown in the face-on view. Each spiral arm is labeled with an id number. The color scale shows the column density distribution but the gas with a density higher than $n_\mathrm{min}=10^{8}\,\mathrm{cm^{-3}}$ is shown. 
}
\label{fig_kinematic_sim}
\end{figure*}

Note that Fig.~\ref{fig_kinematic}a displays the line-of-sight velocity, which is a composite of the infall velocity and the rotation velocity. In contrast, Fig.~\ref{fig_kinematic_sim}a shows the infall velocity and rotation velocity separately along the arms, based on the simulation with 3D velocity components. The location of the arms in the simulation is depicted in Figure~\ref{fig_kinematic_sim}b. Arm 1 (represented by the red lines in Fig.~\ref{fig_kinematic_sim}a) has a considerable infall velocity, which is comparable to, but less than, the rotation velocity in the range of $150~\mathrm{au} \lesssim R \lesssim 300~\mathrm{au}$. Arms 2 and 3 (blue and green lines) exhibit considerably larger infall velocity than rotation velocity in the outer part where $R \gtrsim 100$~au. Therefore, Arms 1, 2, and 3 are classified as infalling streamers, but their infall velocities are less than the freefall velocity (black solid line). In the inner region where $R \lesssim 100$~au, all the arms exhibit rotation velocities that exceed the infall velocities and the Keplerian velocity (black dashed line) due to the gravitational torque from the orbiting protostars.


\subsection{Multiple Formation and Magnetic Field Strength}

\begin{figure*}
\includegraphics[width=\textwidth]{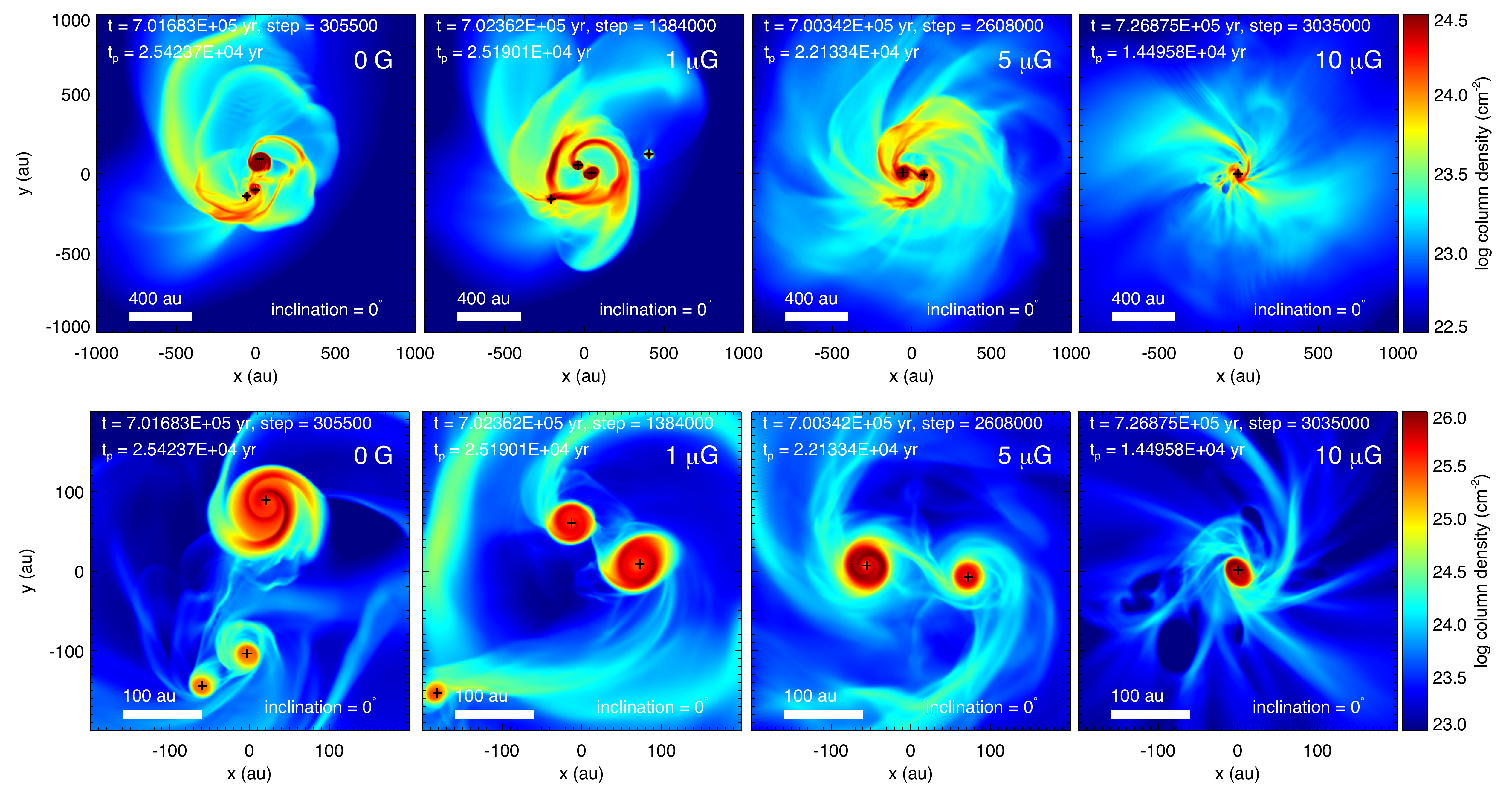}
	\caption{
Comparison of the surface density distributions among the HD/MHD models with initial magnetic field strengths of $0$\,G, $1\,\mu$G, $5\,\mu$G, and $10,\mu$G, from left to right. Upper panels and lower panels display the $(1000\,\mathrm{au})^2$ and $(200\,\mathrm{au})^2$ regions, respectively. The models produce triple stars (0\,G model), quadruple stars ($1,\mu$G model), binary stars ($5\,\mu$G model), and a single star ($10\,\mu$G model). The HD model (0\,G) shown here is calculated using the second-order accuracy scheme, rather than the third-order accuracy scheme, to ensure a fair comparison with the MHD models.
} 
\label{fig_mhdmodels}
\end{figure*}

A magnetic field must be weak to form a multiple stellar system via the fragmentation of a cloud core or a disk \citep{Machida2005}. To test the effect of magnetic field on the formation of multiples and the following development of large spiral arms, we also ran the magneto-hydrodynamic (MHD) simulations varying the initial magnetic field strength and present the results in Fig.~\ref{fig_mhdmodels}. 

The initial condition and numerical models are the same as those of the HD model, but the magnetic fields are included. Uniform magnetic fields perpendicular to the filamentary cloud are imposed as the initial condition because recent observations have suggested that dense filaments tend to have perpendicular magnetic fields \citep{Pattle22}. Three models with the initial magnetic field strengths of 1, 5, and $10~\mu$G are considered. 
The magnetic field strength is often described in terms of the mass-to-flux ratio, and the critical magnetic field strength \citep{Nakano1978,Tomisaka1988} is estimated as $B_\mathrm{cr} = 2 \pi G^{1/2} \Sigma = 20~\mu$G for our models. The initial magnetic field strengths of $1, 5,$ and $10\,\mu$G are therefore 0.05, 0.25, and $0.5 B_\mathrm{cr}$, respectively (corresponding to the non-dimensional mass-to-flux ratios of $\mu=20, 4, 2$). Thus, only magnetically super-critical clouds are considered here.
For an MHD scheme, we adopted Boris-HLLD \citep{Matsumoto2019b}, which allows us to follow a long-term evolution even in the cases of strong magnetic fields. The ohmic dissipation is considered according to the previous simulations \citep{Matsumoto2011,Matsumoto2017}.
The MHD scheme has second-order accuracy in space to maintain numerical stability, while the HD scheme has third-order accuracy. 

\begin{figure*}
\includegraphics[width=0.5 \textwidth]{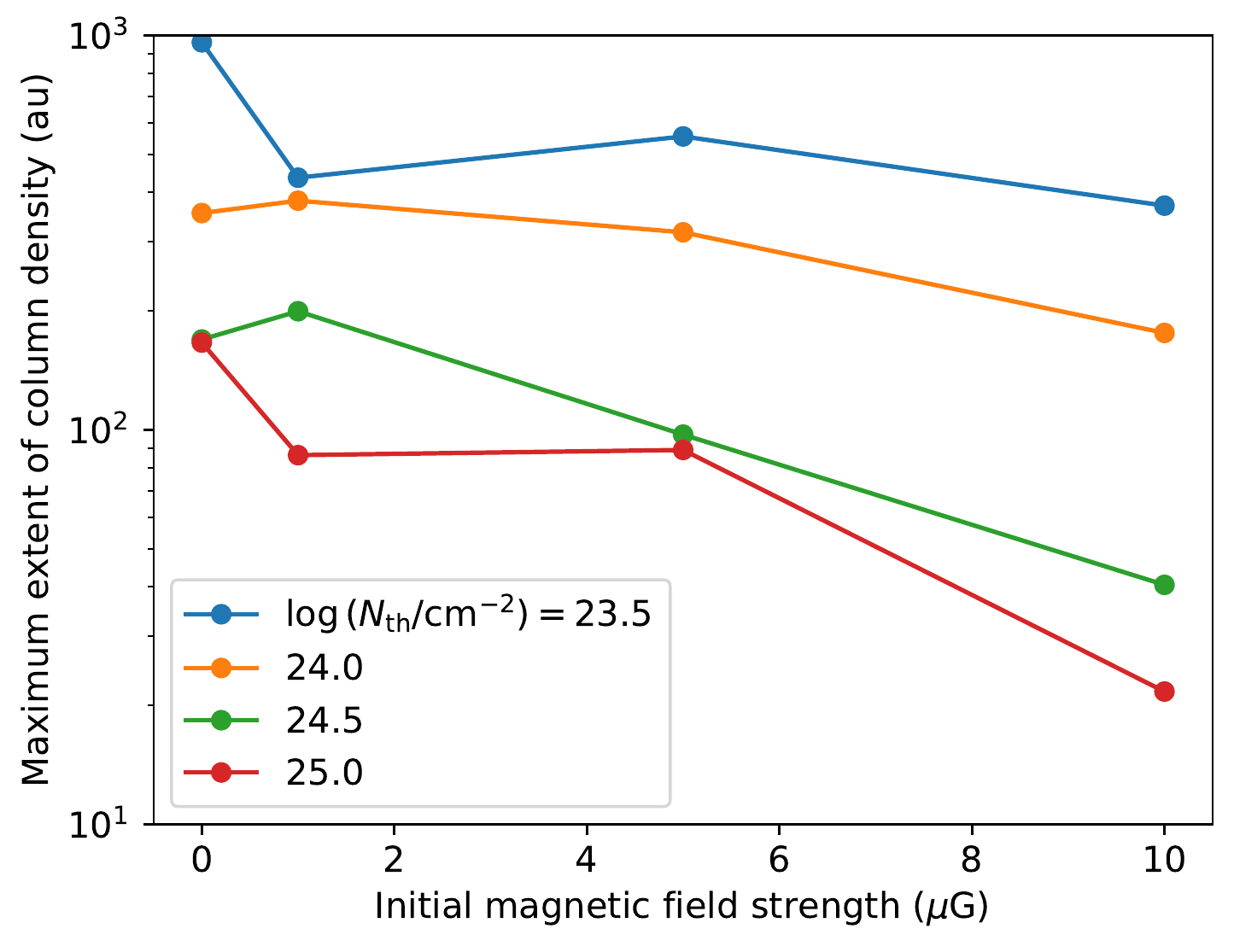}
	\caption{Maximum extent of the column density distribution as a function of initial magnetic field for given thresholds of the column density ($N_\mathrm{th}$) for the models shown in Fig.~\ref{fig_mhdmodels}, indicating the maximum extent of the spiral streamers/arms for each model. The maximum extent is determined by measuring the longest distance between the gas structure with a column density higher than $N_\mathrm{th}$ and the center of mass of the protostars.
} 
\label{fig_armlength}
\end{figure*}

The MHD models exhibit similar evolution to the HD model; the filamentary cloud fragments into cloud cores, which undergo gravitational collapse. A multiple or single-star system forms at the center of the cloud core, depending on the initial magnetic field strength. 
Models with weaker magnetic fields tend to form multiple systems with more companions (Fig.~\ref{fig_mhdmodels}) and to have more prominent and more extended arms (Fig.~\ref{fig_armlength}). The models with 0, 1, 5, and $10~\mu$G form the systems consisting of three (triple), four (quadruple), two (binary), and one (single) stars, respectively.

According to our MHD simulations, the natal cloud of IRAS 04239+2436 must have a very weak magnetic field, with a non-dimensional mass-to-flux ratio larger than $\mu = 2-4$ to form a multiple system. Recent observations of the submillimeter dust emission polarization also show that the strong magnetic field in large filamentary structures is weakened down to the magnetically super-critical condition in small core scales \citep{Doi2021}. The present observation is providing more strong observational evidence of a reduced effect of the magnetic field in a star-forming dense core \citep{Eswaraiah21}.

\subsection{Planet Formation in Multiple Protostellar Systems}

Accumulating evidence suggests that the planet-forming environment around multiple systems must be very different, generally more hostile, from that around single stars.
 Continuum observations show that dust disks in binary systems are smaller in size \citep{Zagaria21b} as dust grains drift inward faster in those disks \citep{Zagaria21a} and due to tidal truncation. The disk gas also likely dissipates faster in multiple systems, owing to higher accretion rates \citep{Zagaria22}. Disks in binary systems have not only a limited amount of dust grains but also a limited time to form planets. 
The smaller the separation, the worse the prospects; the dust continuum fluxes also tend to increase with the separation between companions \citep{Zagaria21b}.
These facts about disks are consistent with facts about the outcomes:  the frequency of planets in close ($<$100 au) binary systems is low compared to those in wide binaries or around single stars  \citep{Kraus2016,Fontanive2019,Marzari2019,Ngo2016}.
We now add to the difficulties by finding evidence for interaction between disks and envelope in the
protostellar stage.
In IRAS 04239+2436, which is a close binary/multiple system with very small ($<$10 au) continuum disks, planet formation may have been negatively impacted by the interactions.
For all these reasons, IRAS 04239+2436 is  not a likely place for planet formation.

\section{Conclusions}
\label{sec:conclusion}

Although most stars form in multiple star systems, their formation process is still controversial. In contrast to the single star formation process, dynamical interactions take place not only among the protostars, but also between the central compact objects and the infalling gaseous envelope. Because the accretion history can be lost in a few orbital timescales, it is important to capture the signature of these dynamical interactions by zooming into such a system at the earliest possible formation epoch in order to better understand the formation process of multiple star system. 
According to our high resolution and high sensitivity ALMA observation, IRAS 04239+2436, which has been known as a protostellar binary system, shows clear triple spiral arms surrounding a young potentially triple protostellar system in the SO line emission. The spiral arms traced by shocked SO gas act as accretion flows over 400 au in length from the large-scale envelope to the 50 au scale stellar system. The imaging results are compared with the numerical magneto-hydrodynamic simulations of analogous systems. While both magnetic fields and turbulence are important to shaping the natal molecular cloud, our numerical simulations suggest that the multiple asymmetric arms may be characteristic of multiple protostars forming in regions of weak magnetic fields. 

\begin{acknowledgments}

ALMA is a partnership of ESO (representing its member states), NSF (USA) and NINS (Japan), together with NRC (Canada), and NSC and ASIAA (Taiwan), and KASI (Republic of Korea), in cooperation with the Republic of Chile. 
The Joint ALMA Observatory is operated by ESO, AUI/NRAO and NAOJ.
This paper makes use of the following ALMA data: ADS/JAO.ALMA\#2015.1.01397.S. 
JEL is supported by the National Research Foundation of Korea (NRF) grant funded by the Korea government (MSIT) (grant number 2021R1A2C1011718.
DH is supported by Centre for Informatics and Computation in Astronomy (CICA)  and grant number 110J0353I9 from the Ministry of Education of Taiwan. DH acknowledges support from the Ministry of Science of Technology of Taiwan through grant number 111B3005191. 
Numerical computations were carried out in part on XC50 (ATERUI II) at the Center for Computational Astrophysics (CfCA), National Astronomical Observatory of Japan. TM is supported by JSPS KAKENHI Grant Numbers 23K03464, 18H05437, 17K05394.
This publication also makes use of data products from NEOWISE, which is a project of the Jet Propulsion Laboratory/California Institute of Technology, funded by the Planetary Science Division of the National Aeronautics and Space Administration.

\end{acknowledgments}

%

\vspace{5mm}
\facilities{ALMA}


\software{CASA \citep{McMullin2007}, \texttt{SFUMATO} \citep{Matsumoto2007} }

\bibliography{iras04239_ref}{}
\bibliographystyle{aasjournal}


\clearpage

\clearpage

\clearpage

\clearpage

\clearpage

\clearpage

\clearpage

\clearpage

\clearpage

\clearpage

\clearpage

\clearpage

\clearpage

\clearpage

\end{document}